\documentclass[12pt]{article}
\usepackage{graphicx}
\usepackage{psfrag}
   \usepackage{adjustbox}
\usepackage{amssymb}
\usepackage[affil-it]{authblk}
\usepackage{natbib}
\usepackage{xcolor}
\usepackage{hyperref}
\usepackage{amsmath}
\def \IR{\hbox{{\rm I}\kern-.2em\hbox{{\rm R}}}}
\title{Geostatistical methods for disease mapping and visualization using data from spatio-temporally referenced prevalence surveys}
\author{Emanuele Giorgi$^1$, Peter J. Diggle$^1$, 
Robert W. Snow$^{2,3}$,\\ Abdisalan M. Noor$^{2}$}
\affil{ \small
$^1$ Lancaster Medical School, Lancaster University, Lancaster, UK \\
$^2$ Population and Health Theme, Kenya Medical Research Institute - Wellcome Trust Research Programme, Nairobi, Kenya \\
$^3$ Centre for Tropical Medicine and Global Health, Nuffield Department of Clinical Medicine, University of Oxford, Oxford, UK}

\DeclareMathOperator{\corr}{corr}

\begin{document}

\maketitle

\begin{abstract}
In this paper we set out general principles and develop geostatistical methods for the analysis of data from spatio-temporally referenced prevalence surveys. Our objective is to provide a tutorial guide that can be used in order to identify parsimonious geostatistical models for prevalence mapping. A general variogram-based Monte Carlo procedure is proposed to check the validity of the modelling assumptions. We describe and contrast likelihood-based and Bayesian methods of inference, showing how to account for parameter uncertainty under each of the two paradigms. We also describe extensions of the standard model for disease prevalence that can be used when stationarity of the spatio-temporal covariance function is not supported by the data. We discuss how to define predictive targets and argue that exceedance probabilities provide one of the most effective ways to convey uncertainty in prevalence estimates. We describe statistical software for the visualization of spatio-temporal predictive summaries of prevalence through interactive animations. Finally, we illustrate an application to historical malaria prevalence data from 1334 surveys conducted in Senegal between 1905 and 2014. 
\end{abstract}

\section{Introduction}
\label{sec:intro}
Model-based geostatistics (MBG) \citep{diggle1998} is a sub-branch of spatial statistics that provides methods for inference on a continuous surface using spatially discrete, noisy data. MBG is increasingly being used in disease mapping applications (e.g. \citet{hay2009, gething2012, diggle2016}), with a particular focus on low-resource settings where disease registries are geographically incomplete or non-existent. \par
We consider data obtained by sampling from a set of potential locations within an area of interest $A$, repeatedly at each of a sequence of times $t_{1},\ldots,t_{N}$. At each sampled location, individuals are then tested for the disease under investigation. The data-format can be formally expressed as
\begin{equation}
\label{eq:data_format}
\mathcal{D} = \{(x_{ij}, t_{i}, y_{ij}, n_{ij}): x_{ij} \in A, j=1,\ldots,m_{i}, i=1,\ldots,N\},
\end{equation}
where $x_{ij}$ is the location of the $j$th of $m_i$ sampling units
 at time $t_{i}$, $n_{ij}$ is the number of tested individuals at $x_{ij}$ and $y_{ij}$ is the  number of positively identified cases.  \par

The methodology described in this paper can be equally applied to longitudinal or repeated cross-sectional designs. For this reason, we re-write \eqref{eq:data_format} as
$$
\mathcal{D} = \{(x_{i},t_{i},n_{i},y_{i}): x_{i} \in A, i=1,\ldots,N^*\},
$$ 
where $N^* = \sum_{i=1}^N m_{i}$ and either or both of the $x_{i}$ and $t_{i}$ may include replicated values.\par

An essential feature of the class of problems that we are addressing in this paper is that the
locations $x_i$ are a discrete set of sampled points within a spatially continuous region of interest.
Another possible  format for prevalence data, which we do not consider in the present study, 
is a  small-area data-set.  In this case, locations $x_i$ are reference locations associated with a partition of
$A$ into $n$ sub-regions.  Disease registries in relatively well developed countries
often use this format, both for administrative
convenience and, in associated publications such as health atlases, to preserve individual confidentiality; see, for example,  \citep{lopezabente2007} or
\citep{hansell2014}. In low-resource settings, this is also often the format of data from demographic surveillance systems, such as Demographic and Health Surveys  (\texttt{dhsprogram.com}), which are nationally representative surveys conducted about every five years to collect information on   population, health and nutrition indicators; see, for example, \citet{mercer2015} for an analysis of data of this kind. \par

A geostatistical model for data of the kind specified by  (\ref{eq:data_format}) 
is that, conditionally on a spatio-temporal process $S(x, t)$ and unstructured random effects $Z(x, t)$, the outcomes $Y$ are mutually independent binomial distributions with number of trials $n$ and probability of being a case $p(x, t)$. Using the conventional choice of a logistic link function, although other choices are also available, we can then write
\begin{equation}
\label{eq:st_model}
\log\left\{\frac{p(x_{i},t_{i})}{1-p(x_{i},t_{i})}\right\} = 
 d(x_{i}, t_{i})^\top \beta + S(x_{i}, t_{i})+Z(x_{i}, t_{i}),
\end{equation}
where $d(x_{i}, t_{i})$ is a vector of spatio-temporally referenced explanatory variables with associated regression coefficients $\beta$. The spatio-temporal random effects $S(x_{i}, t_{i})$ can be interpreted as the cumulative effect of unmeasured spatio-temporal risk factors. These are modelled as a Gaussian process with stationary variance $\sigma^2$ and correlation function 
\begin{equation}
\label{eq:st_cor}
\corr\{S(x,t), S(x',t')\} = \rho(x,x',t,t'; \theta),
\end{equation}
where $\theta$ is a vector of parameters that regulate the scale of the spatial and temporal correlation, the strength of space-time interaction and the smoothness of the process $S(x,t)$. Finally, the unstructured random effects $Z(x_{i}, t_{i})$ are assumed to be independent zero-mean Gaussian variables with variance $\tau^2$, to account for extra-binomial variation within a sampling location. In particular applications, this can represent
 non-spatial random variation, such as genetic or behavioural variation between co-located individuals, 
 spatial variation on a scale smaller than the minimum observed distance between sampled locations,
or
a combination of the two.\par

The model \eqref{eq:st_model} can be used to address two related, but different, research questions. \par
\textit{Estimation: what are the risk factors associated with disease prevalence?} In this case the focus of scientific interest is on the regression coefficients $\beta$. \par
\textit{Prediction: how to interpolate the spatio-temporal pattern of disease prevalence?} The scientific focus is, in this case, on $d(x,t)^\top\beta + S(x,t)$ at both sampled and unsampled locations $\mathcal{X}$ and times $\mathcal{T}$. In some cases, the scientific interest may be more narrowly focused on $S(x,t)$, in order to identify areas of relatively low and high spatio-temporal variation that is not explained by the available explanatory variables. \par

Modelling of the residual spatio-temporal correlation through $S(x,t)$ is crucial in both cases: in the first case, in order to deliver valid inferences on the regression relationships by accurately quantifying the uncertainty in the estimate of $\beta$ \citep{thomson1999}; in the second case, to borrow strength of information across observations $y_{i}$ by exploiting their spatial and temporal correlation.  \par
The use of explanatory variables $d(x, t)$ can also be beneficial in two ways: a simpler model for $S(x,t)$ can be formulated by explaining part of the spatio-temporal variation in prevalence through $d(x,t)$; more precise spatio-temporal predictions between data-locations also result from exploiting the association between disease prevalence and $d(x,t)$. \par
Here, we focus our attention on spatio-temporal prediction of disease prevalence. Our aim is to provide a general framework that can be used as a tutorial guide to address some of the statistical issues common to any spatio-temporal analysis of data from prevalence surveys, especially when sampling is carried out over a large geographical area or time period, or both. More specifically, we provide answers to each of the following research questions.
How can we specify a parsimonious spatio-temporal model while taking account of the main features of the underlying process?
How can we extend model \eqref{eq:st_model} in order to account for non-stationary patterns of prevalence?
What are the predictive targets that we can address using our model for disease prevalence?
How can we effectively visualise the uncertainty in spatio-temporal prevalence estimates?
These issues have only partly been addressed in current spatio-temporal applications of model-based geostatistics for disease prevalence mapping. Some of these are: \citet{clements2006}
on schistosomiasis in Tanzania;
\citet{gething2012} on the world-wide distribution of {\it Plasmodium vivax};
\citet{hay2009} and \citet{noor2014} on the world-wide and Africa-wide distributions of {\it Plasmodium falciparium}; \citet{snow2015} on historical mapping of malaria in the Kenyan Coast area; \citet{bennet2013} on the mapping of malaria transmission intensity in Malawi; \citet{kleinschmidt2001} on malaria incidence in KwaZulu Natal, South Africa; \citet{kleinschmidt2007} on HIV in South Africa; \citet{magalhaes2011} on anemia in preschool-aged children in West Africa;
\citet{raso2005} on schistosomiasis in C\^ote D'Ivoire; \citet{pullan2011} on soil-transmitted infections in Kenya; \citet{zoure2014} on river blindness in the 20 participating countries of the African
Programme for Onchocerciasis control. In almost all of these cases, the adopted spatio-temporal model is only assessed with respect to its predictive performance, using ROC curves and prediction error summaries. In our view, a validation check on the adopted correlation structure in the analysis should precede geostatistical prediction, as misspecification of the spatio-temporal structure of the field $S(x,t)$ can potentially lead to an inaccurate quantification of uncertainty in the prevalence estimates and, therefore, to invalid inferences. In this paper, we describe the different stages of a spatio-temporal geostatistical analysis and provide tools that directly address the issue of specifying a spatio-temporal covariance structure that is compatible with the data. \par

The paper is structured as follows. Section \ref{sec:design} is a review on geostatistical sampling design, where we show how this might affect our analysis of the data. In Section \ref{sec:methods} we describe principles and provide statistical tools for each of the stages of a spatio-temporal geostatistical analysis. In Section \ref{subsec:expl_analysis}, we define the objectives of an exploratory geostatistical analysis and show how to pursue these using the empirical variogram. In Section \ref{subsec:lik_or_bayes}, we outline and contrast likelihood-based and Bayesian methods of inference. In Section \ref{subsec:diagnostic}, we propose a general Monte Carlo procedure based on the empirical variogram, in order to check the validity of the assumed spatio-temporal correlation function for $S(x,t)$. In Sections \ref{subsec:pred_targets} and \ref{subsec:visualization}, we discuss how to define and visualize predictive targets. In Section \ref{sec:app} we illustrate an application to historical mapping of malaria using data from prevalence surveys conducted in Senegal between 1905 and 2014. Section \ref{sec:discussion} is a concluding discussion.

\section{Geostatistical sampling design}
\label{sec:design}
Different design scenarios can give rise to data of the kind expressed by \eqref{eq:data_format}. A good choice of
design depends both on the objectives of the study and on practical constraints. 

In a longitudinal design, data are collected repeatedly over time from the same set of sampled locations. This is an appropriate strategy when temporal variation in the outcome of primary interest
dominates spatial variation, and more obviously
 when the scientific goal is to understand change over time at a set of sentinel locations. A longitudinal design is also cost-effective when
setting up a sampling location is expensive but subsequent data-collection is cheap. 

In a repeated cross-sectional design, a different set of locations is chosen on each sampling
occasion. This sacrifices direct information on changes in disease prevalence over time in favour of more complete spatial coverage. Repeated cross-sectional designs can also be
adaptive, meaning that on any sampling occasion, the choice of sampling locations is informed
by an analysis of the data collected on earlier occasions. Adaptive repeated cross-sectional designs are therefore particularly suitable for applications in which temporal variation either is dominated by spatial variation or can be well explained by available covariates; see \citet{chipeta2016} and \citet{kabaghe2017}. \par

To explain how the sampling design might affect our geostatistical analysis of the data, 
let
 $\mathcal{X} = \{x_{i} \in A : i=1,\ldots,n\}$ 
denote the set of sampling locations arising from the sampling design,  $\mathcal{S} = \{S(x) : x \in A \}$ 
the signal process and  $\mathcal{Y} = \{Y_{i} : 1=1,\ldots,n\}$ the outcome data. \par

A sampling design is deterministic if it consists of a set of pre-defined sampling locations, and  
 stochastic if the locations are a probability-based selection from a  set of candidate
designs. In the latter case $\mathcal{X}$ is a finite point process on the region of interest $A$. Let $[\cdot]$ denote 
 ``the distribution of.'' Our model for the outcome
data is then obtained by integrating out $\mathcal{S}$ from the joint 
distribution $[\mathcal{X}, \mathcal{S}, \mathcal{Y}]$, i.e.
\begin{equation}
\label{eq:data_model}
[\mathcal{X},\mathcal{Y}] = \int [\mathcal{X}, \mathcal{S}, \mathcal{Y}] \: d\mathcal{S}.
\end{equation}
From a modelling perspective, the most natural factorization of the integrand in the above equation is as
\begin{equation}
\label{eq:joint}
[\mathcal{X}, \mathcal{S}, \mathcal{Y}] = [\mathcal{S}][\mathcal{X} | \mathcal{S}][\mathcal{Y} | \mathcal{X}, \mathcal{S}].
\end{equation}
The design is {\it non-preferential}
 if $[\mathcal{X} | \mathcal{S}] = [\mathcal{X}]$,  in which case \eqref{eq:data_model} becomes
\begin{equation}
\label{eq:data_model_non_pref}
[\mathcal{X},\mathcal{Y}] = [\mathcal{X}]\int [\mathcal{S}][\mathcal{Y} | \mathcal{X}, \mathcal{S}] \: d\mathcal{S}.
\end{equation}
Hence, under non-preferential sampling schemes, inference about ${\cal S}$ and/or ${\cal Y}$
 can be conducted legitimately by simply conditioning on the observed set of locations, ${\cal X}$. 

The simplest example of a probabilistic sampling design is completely random sampling. This can be interpreted, according to context, either as a random sample from a finite, pre-specified set of potential sampling locations or as an independent random sample from the continuous uniform distribution
on $A$. Other examples include spatially stratified random sampling designs, which
 consist of a collection of completely random designs, one in each of a number of subdivisions of $A$, and systematic sampling designs, in which the sampled locations
form a regular (typically rectangular) lattice to cover $A$, strictly with the first lattice-point
chosen at random, although in practice this is often ignored.

Here as in other areas of statistics, the choice of  sampling design affects inferential precision.
  If, for example,  the inferential target is the underlying
spatially continuous prevalence surface, $p(x,t^*)$ at a future time $t^*$, a possible design 
goal for geostatistical prediction would be to minimise the spatial average of 
the mean squared error,
$$\int_A {\rm E}[\{\hat{p}(x,t^*) - p(x,t^*)\}^2] dx,$$
where $\hat{p}(x, t^*)$ is a predictor for $p(x,t^*)$ obtained from \eqref{eq:st_model}. In contrast, a possible design goal for estimation of the
relationship between a covariate $d(x,t)$ and disease prevalence would be to minimise the variance of 
the estimated regression parameter, $\hat{\beta}$. \par

Efficient sampling designs for 
spatial prediction generally require
sampled locations to be distributed more evenly over $A$ than would result from completely
random or stratified random sampling;
see, for example,  \cite{matern1986}. \par

Stratified sampling often provides a more cost-effective design than simple random sampling from the general population. In cases where the strata correspond to sub-populations associated with different disease risk levels, a geostatistical model should account for the stratification
 through the use of an appropriate explanatory variable. To illustrate this, consider, for example, a population consisting of $K$ strata which correspond to a partition of the region of interest, $A$, into non-overlapping regions $\mathcal{R}_{k}$ for $k=1,\ldots,K$. We then take a random sample from each region $\mathcal{R}_{k}$ so that each location $x \in \mathcal{R}_{k}$ has probability of being selected proportional to the population of $\mathcal{R}_{k}$. If it is known that each of the strata $\mathcal{R}_{k}$ is associated with different levels in disease risk, this can be accounted for by including a factor variable in \eqref{eq:st_model} with $K-1$ levels or, if $K$ is large, using random effects at stratum-level. In some cases the strata can also be grouped into sub-populations which are known to differ in their exposure to the disease. For example, let us assume that each stratum can be classified as being urban or rural and that these two types of areas are associated with different risk levels, i.e.
\begin{equation}
\label{eq:strat_sampling}
\log\left\{\frac{p(x_{i},t_{i})}{1-p(x_{i},t_{i})}\right\} = 
 \beta + \alpha u(x_{i})+ S(x_{i}, t_{i})+Z(x_{i}, t_{i}),
\end{equation}
where $u(x_{i})$ is an indicator function that takes value 1 if $x_{i} \in \mathcal{R}_{k}$ and $\mathcal{R}_{k}$ is urban, and 0 otherwise. Under this model, it follows that 
$$
[\mathcal{Y}, \mathcal{S}, \mathcal{X}] = [\mathcal{X}] [\mathcal{S}] [\mathcal{Y} | \mathcal{S}, \mathcal{X}]
$$
hence \eqref{eq:strat_sampling} does not constitute an instance of preferential sampling. This shows that variables used in the design should be included in the model when these are associated with the outcome of interest so as to ensure that the sampling is non-preferential. For a wider discussion on this issue in the context of standard regression models, we refer to \citet{skinner2017} and \citet{lumley2017}.\par

Another common design in practice is the opportunistic
sampling design \citep{hedt2011}, in which data are
collected at convenient places, for example from presentations at health clinics, a market or a school. The
limitations of this are obvious: opportunistic samples may not 
be representative of the target population and so not deliver unbiased estimates of $p(x,t)$.  Also, as unmeasured factors relating to the disease
in question are likely to affect an indivudual's decision to present,
 the assumption of non-preferential sampling is questionable.
 For example, areas with atypically
high or low levels of $p(x,t)$ may have been systematically oversampled; see \citet{diggle2010} and \citet{pati2011} for a discussion and formal solution to the problem of geostatistical inference under preferential sampling. \par 

\citet{giorgi2015} address
 the issue of combining data from multiple prevalence surveys, with a mix of random and opportunistic sampling designs. By developing a multivariate geostatistical model that enables estimation of the bias from opportunistic samples, they show that combining information from multiple studies  can lead to more precise estimates of prevalence, provided that at least one of these is known to be unbiased. \par

In the remainder of this paper, we shall focus our attention on the case of prevalence data obtained from a non-preferential sampling design.

\section{Methods}
\label{sec:methods}
\begin{figure}
\begin{center}
\includegraphics[scale=0.9]{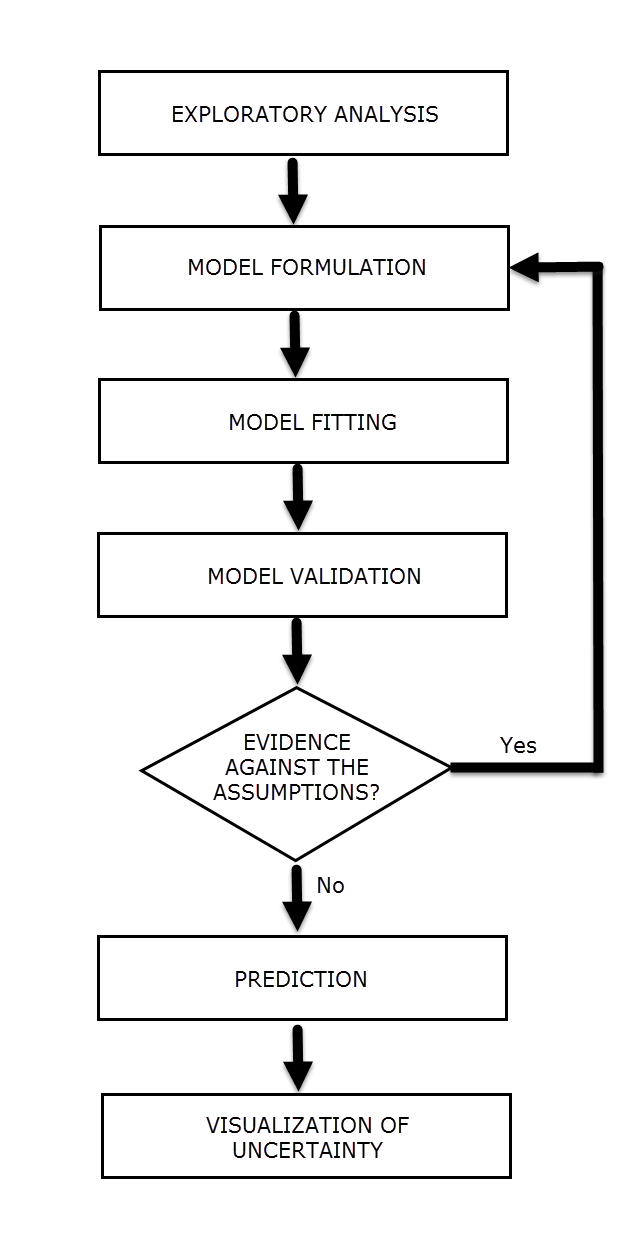}
\caption{Diagram of the different stages of a statistical analysis. \label{fig:diag_stat_analysis}}
\end{center}
\end{figure}
In this Section we provide a general framework for the analysis of data from spatio-temporally referenced prevalence surveys. Figure \ref{fig:diag_stat_analysis} shows the different stages of the analysis as a cycle that terminates when all the modelling assumptions are supported by the data. In our context, visualization of the results also plays an important role in order to display the spatio-temporal patterns of estimated prevalence and to communicate uncertainty effectively.

\subsection{Exploratory analysis: the spatio-temporal variogram}
\label{subsec:expl_analysis}
The usual starting point for a spatio-temporal analysis of prevalence data is an analysis based on a binomial mixed model without spatial random effects, i.e. $S(x,t)=0$ for all $x$ and $t$. Let $\tilde{Z}(x_{i},t_{i})$ denote a point estimate, such as the predictive mean or mode, of the unstructured random effects $Z(x_{i}, t_{i})$ from the non-spatial binomial mixed model. We then analyse  $\tilde{Z}(x_{i},t_{i})$ to pursue the two following objectives:

\begin{enumerate}
\item testing for presence of residual spatio-temporal correlation;
\item formulating a model for \eqref{eq:st_cor} and providing an initial guess for $\theta$.
\end{enumerate}

We make a working assumption that $S(x,t)$ is a stationary and isotropic process, hence 
\begin{equation}
\label{eq:iso_st_cor}
\rho(x,x',t,t'; \theta) = \rho(u,v; \theta), 
\end{equation}
where $u = \|x-x'\|$, with $\|\cdot\|$ denoting the Euclidean distance, and $v = |t-t'|$.  \par
The {\it variogram} can then be used to formulate and validate models for the spatio-temporal correlation in \eqref{eq:st_cor}. Let $W(x,t) = S(x,t)+Z(x,t)$, where $S(x,t)$ and $Z(x,t)$ are specified as in \eqref{eq:st_model}; the spatio-temporal variogram of this process is given by
\begin{eqnarray}
\label{eq:variog}
\gamma(u,v; \theta) = \frac{1}{2}E[\{W(x,t)-W(x',t')\}^2] = \tau^2+\sigma^2[1-\rho(u,v; \theta)].
\end{eqnarray}
\par
We refer to this as the \textit{theoretical} variogram, since it is directly derived from the theoretical model for the process $W(x,t)$. \par

We use $\tilde{Z}(x_i, t_i)$ to estimate the unexplained extra-binomial variation in prevalence, at observed locations $x_{i}$ and times $t_{i}$. Let $n(u,v)$ denote the pairs $(i,j)$ such that $\|x_{i}-x_{j}\|=u$ and $|t_{i}-t_{j}|=v$; the \textit{empirical} variogram is then defined as 
\begin{equation}
\label{eq:empir_vario}
\tilde{\gamma}(u,v) = \frac{1}{2|n(u,v)|}\sum_{(i,j) \in n(u,v)} \{\tilde{Z}(x_{i},t_{i})-\tilde{Z}(x_{j},t_{j})\}^2,
\end{equation}
where $|n(u,v)|$ is the number of pairs in the set. \par
Testing for the presence of residual spatio-temporal correlation can be carried out using the following Monte-Carlo procedure:

\begin{itemize}
\item[(Step 1)] permute the order of the data, including $\tilde{Z}(x_{i}, t_{i})$, while holding $(x_{i}, t_{i})$ fixed;
\item[(Step 2)] compute the empirical variogram for $\tilde{Z}(x_{i}, t_{i})$;
\item[(Step 3)] repeat($i$) and ($ii$) a large enough number of times, say $B$;
\item[(Step 4)] use the resulting $B$ empirical variograms to generate $95\%$ tolerance intervals at each of the pre-defined distance bins.
\end{itemize}
If $\tilde{\gamma}(u,v)$ lies outside these intervals, then the data show evidence of residual spatio-temporal correlation. If this is the case, the next step is to specify a functional form for $\rho(u,v)$. \par

\citet{gneiting2002} proposed the following class of spatio-temporal correlation functions
\begin{equation}
\label{eq:gneiting_cor1}
\rho(u,v;\theta) = \frac{1}{(1+v/\psi)^{\delta+1}}
\exp\left\{-\frac{u/\phi}{(1+v/\psi)^{\xi/2}}\right\}, 
\end{equation}
where $\phi$ and $(\delta,\psi)$ are positive parameters that  determine the rate at which the spatial and temporal correlations decay, respectively.  When $\xi=0$ in (\ref{eq:gneiting_cor1}),
$\rho(u,v;\theta) = \rho_1(u) \rho_2(v)$ where $\rho_1(\cdot)$ and $\rho_2(\cdot)$ are purely spatial and
purely temporal correlation functions, respectively. Any spatio-temporial correlation function
that factorises in this way is called {\it separable}.
In this sense,  the parameter $\xi \in [0,1]$ represents the extent of non-separability. \citet{stein2005} provides a detailed analysis of the properties of space-time covariance functions and highlights the limitations of using separable families. However, fitting of complex space-time covariance models requires
 more data than, in our experience,  is typically available in prevalence mapping applications.
In the application of Section \ref{sec:app}, we show that only $\psi$ and $\phi$ in \eqref{eq:gneiting_cor1} can be estimated with an acceptable level of precision, whilst the data are poorly informative with respect to the other covariance parameters, in which case the parsimony principle  favours a separable model. Note, incidentally, that
separability is implied by, but does not imply, that $S(x,t)$ can be factorised as $S_1(x) S_2(t)$, which would be
a highly artificial construction. 
\par

A spatio-temporal correlation function is separable if 
$$
\rho(u,v;\theta) = \rho_{1}(u;\theta_{1})\rho_2(v;\theta_{2}),
$$
where $\theta_{1}$ and $\theta_{2}$ parametrise the purely spatial and temporal correlation functions, respectively; in the case of \eqref{eq:gneiting_cor1}, this is separable when $\xi=0$. Separable correlation functions are computationally convenient when joint predictions of prevalence are required at different time points over the same set of prediction locations. Checking the validity of the separability assumption can be carried out using the likelihood-ratio test for models such as \eqref{eq:gneiting_cor1}, where separability can be recovered as a special case. \par
Once a parametric model has been specified, an initial guess for $\theta$ can be used to initialise the maximization of the likelihood function.
One way to obtain an initial guess is to choose the value of $\theta$ that minimizes the sum of squared differences between the theoretical and empirical variogram ordinates. Section 5.3 of \citet{diggle2007book} describes the least squares algorithm and other, more refined methods to fit a parametric variogram model to an empirical variogram. However, in our view, variogram-based techniques should only be used for exploratory analysis and diagnostic checking. For parameter estimation and formal inference, likelihood-based and Bayesian methods are more efficient and more objective.

\subsection{Parameter estimation and spatial prediction}
\label{subsec:lik_or_bayes}
We now outline likelihood-based and Bayesian methods of parameter estimation for the model in \eqref{eq:st_model}. 

\subsubsection{Likelihood-based inference}
\label{subsubsec:likelihood}
Let $\lambda^\top=(\beta^\top,\sigma^2,\theta^\top)$ denote the set of unknown model parameters, including regression coefficients $\beta$, the variance $\sigma^2$ of $S(x,t)$ and covariance parameters $\theta$. We use $[\cdot]$ as a shorthand notation for ``the distribution of''. The likelihood function is then obtained from the marginal distribution of the outcome $y^\top=(y_{1},\ldots,y_{n})$ by integrating out the random effects $W^\top=(W(x_{1},t_{1}),\ldots,W(x_{n}, t_{n}))$ to give
\begin{equation}
\label{eq:lik}
L(\lambda) = [y|\lambda] = \int [W, y | \lambda] \: dW.
\end{equation}
In general, the integral in \eqref{eq:lik} is intractable. However, numerical integration techniques or Monte Carlo methods can be used for approximate evaluation and maximization of the likelihood function, as required for classical inference \citep{geyer1992, geyer1994, geyer1996, geyer1999}. See \citet{christensen2004} for a detailed description of the Monte Carlo maximum likelihood estimation method in a geostatistical context. \par
In our application of Section \ref{sec:app}, we use the following approach to approximate \eqref{eq:lik}. Let $\lambda_{0}$ represent our best guess of $\lambda$. We then rewrite \eqref{eq:lik} as
\begin{eqnarray}
\label{eq:lik0}
L(\lambda) &=& \int \frac{[W, y | \lambda]}{[W, y | \lambda_0]} [W, y | \lambda_0]\: dW \nonumber \\
&\propto& \int \frac{[W, y | \lambda]}{[W, y | \lambda_0]} [W | y, \lambda_0]\: dW \nonumber \\
&=& E\left\{ \frac{[W, y | \lambda]}{[W, y | \lambda_0]}\right\},
\end{eqnarray}
where the expectation in the above equation is taken with respect to $[W | y, \lambda_0]$. Using MCMC algorithms, we then generates $B$ samples from $[W | y, \lambda_0]$, say $w_{(i)}$, and approximate \eqref{eq:lik0} as
$$
L_{B}(\lambda) = \frac{1}{B} \sum_{i=1}^B \frac{[w_{(i)} | y, \lambda]}{[w_{(i)} | y, \lambda_0]}.
$$
We maximize $L_{B}(\lambda)$ using a Broyden-Fletcher-Goldfarb-Shanno algorithm \citep{fletcher1987}, which incorporates analytical expressions for the first and second derivatives of $L_{B}(\lambda)$. Let $\hat{\lambda}_{B}$ denote the Monte Carlo maximum likelihood estimate of $\lambda$. We then set $\lambda_0 = \hat{\lambda}_{B}$ and repeat the outlined procedure until convergence. \par

To simulate from $[W | y, \lambda_0]$, we first reparametrise the model based on $\tilde{W} = \hat{\Sigma}^{-1/2}(W-\hat{w})$, where $\hat{w}$ is the mode of $[W | y, \lambda_0]$ and $\hat{\Sigma}$ is the inverse of the negative Hessian of $[W | y, \lambda_0]$ at the mode $\hat{w}$. At each iteration of the MCMC, we propose a new value for $\tilde{W}$, given the current value $w$, using a Langevin-Hastings algorithm with a Gaussian proposal distribution having mean 
$$
w + (h/2) \nabla \log[w | y, \lambda_0] 
$$
and covariance matrix given by $hI$, where $I$ is the identity matrix and $h$ is tuned so that the acceptance rate is 0.574 \citep{roberts1998}. \par

Other approaches that have been proposed to maximize \eqref{eq:lik} are based on the expectation-maximization algorithm \citep{zhang2002} and the Laplace approximation \citep{bonat2016}. \par

Let $W^*$ denote the vector of values of $W(x,t)$ at a set of unobserved times and locations. The formal solution to the prediction problem is to evaluate
the conditional distribution of $W^*$ given the data $y$. Although the
joint predictive distribution of the elements of $W^*$
is intractable, it is possible to simulate samples from this distribution. \par
 If we assume,
unrealistically, that $\lambda$ is known, the predictive distribution of $W^*$ is given by
\begin{equation}
[W^* | y, \lambda] = \int [W^*, W | y, \lambda] \: dW = \int [W | y, \lambda] [W^* | W, y, \lambda] \: dW = \int [W | y, \lambda] [W^* | W, \lambda] \: dW.
\label{eq:predictive_known}
\end{equation}
See Chapter 4 of \citet{diggle2007book} for explicit expressions. \par
If,  more realistically, $\lambda$ is unknown, {\it plug-in} prediction consists of replacing
$\lambda$ in (\ref{eq:predictive_known}) by an estimate $\hat{\lambda}$, preferably the maximum
likelihood estimate. A legitimate criticism
of this is that the resulting predictive probabilities ignore the inherent uncertainty in
$\hat{\lambda}$. However, this can be taken into account within a likelihood-based inferential framework as follows. Let 
$\hat{\Lambda}$ denote the maximum likelihood estimator of $\lambda$. 
We define the predictive distribution of $W^*$ as
\begin{equation}
[W^*| y] = \int \int [\hat{\Lambda}] [W | y, \hat{\Lambda}] [W^* | W, \hat{\Lambda}] \: dW \: d\hat{\Lambda},
\label{eq:predictive_Lik}
\end{equation}
where $[\hat{\Lambda}]$ denotes the sampling distribution  of  the maximum likelihood 
estimator $\hat{\Lambda}$. Equation (\ref{eq:predictive_Lik})  acknowledges the
uncertainty in $\hat{\Lambda}$ by expressing the predictive distribution 
$[W^*| y]$ as the expectation
  of the plug-in predictive distribution (\ref{eq:predictive_known})
with respect to the sampling distribution of $\hat{\Lambda}$.
This can then be approximated using a multivariate Gaussian distribution with
 mean given by the observed MLE, $\hat{\lambda}$, and covariance matrix given by
$$
\left[-\frac{\partial^2 \log L(\hat{\lambda})}{\partial^2 \lambda}\right]^{-1}.
$$
In our experience, the quality of the Gaussian approximation
is improved considerably by applying a log-transformation to each of the covariance parameters. If the Gaussian approximation remains questionable, 
 a more computationally intensive alternative is a parametric bootstrap consisting of the following steps: simulate a number of binomial data-sets using the plug-in MLE for $\lambda$; for each simulated data-set, carry out parameter estimation by maximum likelihood. The resulting set of bootstrap estimates for $\lambda$ can then be used to approximate the distribution of $\hat{\Lambda}$. We give an example of these approaches in the case-study of Section \ref{sec:app}.

\subsubsection{Bayesian inference}
\label{subsubsec:bayes}
In Bayesian inference, $\lambda$ is treated as a random variable and must be assigned a
prior distribution, $[\lambda]$. Parameter estimation is then carried out through
 the posterior distribution of 
$\lambda$,  which is obtained using Bayes' theorem as
\begin{equation}
\label{eq:bayes}
[\lambda|y] =  \frac{[\lambda][y | \lambda]}{[y]} = \frac{[\lambda]L(\lambda)}{[y]}.
\end{equation}
All other things being equal,
as the sample size increases $L(\lambda)$ becomes more concentrated around the true value of $\lambda$,
the impact of the prior is reduced  and the difference between likelihood-based and Bayesian parameter estimation
 becomes less important. MCMC algorithms can be used for approximate computation of the posterior in \eqref{eq:bayes}. For the Bayesian analysis in the application of Section \ref{sec:app}, we develop an MCMC algorithm which separately updates $\beta$, $\sigma^2$, $\theta$ and $W$. 
Specifically, we use a Metropolis-Hastings algorithm
to update $\log\{\sigma^2\}$ and $\log\{\theta\}$, and a Gibbs sampler 
to update $\beta$. To update the random effect $W$, we use a Hamiltonian Monte Carlo procedure \citep{neal2011}. More computational details on this approach can be found in Section 2.2 of \citet{giorgi2017}. \par
Non-stochastic analytical approximations of \eqref{eq:bayes} can also be obtained using, for example by the use of
 integrated nested Laplace approximations \citep{rue2009}. However, their accuracy should be considered carefully in each spefici context. \citet{joe2008} shows that for binomial mixed models, the smaller the denominator the less accurate is the Laplace approximation. \citet{fong2010}, in a
review of computational methods for Bayesian inference in generalized linear mixed models,
also report poor performance of the INLA method in the case binary responses
  \par

Bayesian predictive inference about $W^*$
 uses a second application of Bayes' theorem to give the predictive 
distribution 
\begin{equation}
[W^*| y] = \int \int [\lambda | y] [W | y, \lambda] [W^* | W, \lambda] \: dW \: d\lambda,
\label{eq:predictive_Bayes}
\end{equation}
where $[\lambda|y]$ is the posterior distribution of $\theta$. Comparison of (\ref{eq:predictive_Bayes}) 
and (\ref{eq:predictive_Lik}) shows that both
are weighted averages of plug-in predictive distributions. The difference between them is that
(\ref{eq:predictive_Bayes}) uses the posterior $ [\lambda | y]$ as the weighting distribution whilst (\ref{eq:predictive_Lik})
uses the sampling distribution $[\hat{\Lambda}]$.
In either case, the weights concentrate increasingly around the maximum likelihood estimate of $\lambda$ as
the sample size increases. \par

In our experience the difference between plug-in prediction using the maximum likelihood estimate $\hat{\lambda}$
and  weighted average prediction
is  often negligible, because the uncertainty in $W^*$  dominates that in $\lambda$.
An intuitive explanation for this is that for
estimation of $\lambda$ all of the
data contribute information,  whereas for prediction of $W(x,t)$ only data at locations and times relatively
close to $x$ and $t$ contribute materially. However, this is not guaranteed, especially when the predictive target is a non-linear property of $W^*$; see, for example, Figure 9a of \citet{diggle2002}. \par

\subsection{Diagnostics and novel extensions}
\label{subsec:diagnostic}

In order to check the validity of the chosen spatio-temporal covariance function, we modify the Monte Carlo algorithm introduced in Section \ref{subsec:expl_analysis} by replacing (Step 1) with following.

\begin{itemize}
\item[(Step 1)] Simulate $W(x_{i},t_{i})$ at observed locations $x_{i}$ and times $t_{i}$, for $i=1,\ldots,n$, from its marginal multivariate distribution under the assumed model. Conditionally on the simulated values of $W(x_{i}, t_{i})$, simulate binomial data $y_{i}$ from \eqref{eq:st_model}. Finally, compute the point estimates $\tilde{Z}(x_{i}, t_{i})$ using the simulated data.
\end{itemize}

In this case, the resulting 95$\%$ tolerance band is generated under the assumption that the true covariance function for $S(x,t)$ exactly corresponds to the one adopted for the analysis. If $\tilde{\gamma}(u,v)$ lies outside the intervals, then this indicates that the fitted covariance function is not compatible with the data. To formally test this hypothesis, we can also use the following test statistic
\begin{equation}
\label{eq:test_statistic}
T = \sum_{k=1}^K |n(u_{k}, t_{k})| [\tilde{\gamma}(u_{k}, v_{k}) - \gamma(u_{k}, v_{k}; \theta)]^2,
\end{equation}
where $u_k$ and $v_k$ are the distance and time separations of the variograms bins, respectively, the $n(u_{k}, t_{k})$ are the numbers of pairs of observations contributing to each bin and $\theta$ is the true parameter value of the covariance parameters. Since $\theta$ is almost always unknown, it can be estimated using either maximum likelihood or Bayesian methods, in which case \eqref{eq:test_statistic} should be averaged over the posterior distribution of $\theta$ using posterior samples $\theta_{(h)}$, i.e.
\begin{equation}
\label{eq:test_statistic_bayes}
T = \frac{1}{B}\sum_{h=1}^B \sum_{k=1}^K |n(u_{k}, t_{k})| [\tilde{\gamma}(u_{k}, v_{k}) - \gamma(u_{k}, v_{k}; \theta_{(h)})]^2.
\end{equation}
The null distribution of $T$ can be obtained using the simulated values for $\tilde{Z}(x_{i}, t_{i})$ from the modified (Step 1) introduced in this section. Let $T_{(h)}$ denote the $h$-th sample from the null distribution of $T$, for $h=1,\ldots,B$. Since evidence against the adopted covariance model arises from large values of $T$, an approximate p-value can be  computed as
$$
\frac{1}{B} \sum_{h=1}^B I[T_{(h)} > t],
$$ 
where $I(a > b)$ takes value 1 if $a > b$ and 0 otherwise, and $t$ is the value of the test statistic obtained from the data. \par

An unsatisfactory result from this diagnostic check could indicate a need for either or both of two extensions to the model: a more flexible family of stationary covariance structures; or non-stationarity induced by parameter variation over time, space or both. \par
In the former case, we note that the correlation function in \eqref{eq:gneiting_cor1} can also be obtained a special case of 
\begin{equation}
\label{eq:gneiting_cor2}
\rho(u,v;\theta) = \frac{1}{(1+v/\psi)^{\delta+1}} \mathcal{M}\left(\frac{u}{(1+v/\psi)^{\xi/2}}; \phi, \kappa \right)
\end{equation}
where $\mathcal{M}\left(\cdot; \phi, \kappa \right)$ is the \citet{matern1986} correlation function with scale and smoothness parameters $\phi$ and $\kappa$, respectively \citep{gneiting2002}. Equation \eqref{eq:gneiting_cor1} is recovered for $\kappa=1/2$. However, the additional parameter introduced, $\kappa$, is likely to be poorly identified. A pragmatic response is to discretise the smoothness parameter $\kappa$ in \eqref{eq:gneiting_cor2} to a finite set of values, e.g. $\{1/2,3/2,5/2\}$, over which the likelihood function is maximized. \par
In the second case, the context of the analysis can provide some insights on the nature of the non-stationary behaviour of the process being studied. For example, if data are sampled over a large geographical area, such as a continent, one may expect the properties of the process $S(x,t)$ to vary across countries. This can then be assessed by fitting the model separately for each country. A close inspection of the parameter estimates for $\theta$ might then reveal which of its components show the strongest variation. Furthermore, if these estimates also show spatial clustering, the vector $\theta$, or some of its components, can be modelled as an additional spatial process, say $\Theta(x)$. The process $S(x,t)$ is then modelled as a stationary Gaussian process conditionally on $\Theta(x)$. A similar argument can also be developed if data are collected over a large time period in a geographically restricted area. In this case, $\theta$ may primarily vary across time and, therefore, could be modelled as a temporal stochastic process. 

\subsubsection{Example: a model for disease prevalence with temporally varying variance}
\label{subsubsec:temp_var}
We now give an example of how model \eqref{eq:st_model} can be extended in order to allow the nature of the spatial variation in disease prevalence to change over time. We replace the spatio-temporal random effect $S(x,t)$ in the linear predictor with
\begin{equation}
\label{eq:temp_var_resid}
S^*(x,t) = B(t) S(x,t),
\end{equation}
where $B^2(t)$ represents the temporally varying variance of $S^*(x,t)$. We then model $\log\{B^2(t)\}$ as a stationary Gaussian process, independent of $S(x,t)$, with mean $-\eta^2/2$, variance $\eta^2$ and one-dimensional correlation function $\rho_{B}(\cdot; \theta_{B})$, with covariance parameters $\theta_{B}$. Note that, using this parametrisation,  $E[B^2(t)]=1$ and, therefore, $V[S^*(x,t)]=\sigma^2$. The resulting process $S^*(x,t)$ is a non-Gaussian process with heavier tails than $S(x,t)$ and correlation function
\begin{eqnarray}
\label{eq:non_gauss_cor}
\corr\{S^*(x,t), S^*(x',t')\} &=& 
\exp\{\eta^2(\rho_{B}(v; \theta_{B})-1)\}\rho(u,v; \theta).
\end{eqnarray}
The likelihood function is obtained as in \eqref{eq:lik} but now with $W(x_{i}, t_{i}) = S^*(x_{i}, t_{i}) + Z(x_{i}, t_{i})$. 
 
\subsection{Defining targets for prediction}
\label{subsec:pred_targets}
Let $\mathcal{P}(W^*) = \{p(x,t): x \in A, t \in [T_{1}, T_{2}]\}$ denote the set of prevalence surfaces covering the region of interest $A$ and spanning the time period $[T_{1}, T_{2}]$. 
Prediction of $\mathcal{P}$ is carried out by first simulating samples from the the predictive distribution of $W^*$, i.e. the
distribution of $W^*$ conditional on the data $y$. From each simulated sample
of $W^*$, we then calculate any required summary, ${\cal T}$ say, of the corresponding
$\mathcal{P}(W^*)$, for example means or selected quantiles
 at any $(x,t)$ of interest. By construction, this generates a sample from the
predictive distribution of ${\cal T}$. Computational details and explicit expressions can be found in \citet{giorgi2017}. \par

Two ways to display uncertainty in the estimates of prevalence are through quantile or exceedance probability surfaces. We define the \textit{$\alpha$-quantile surface} as
\begin{equation}
\label{eq:quantile_surf}
\mathcal{Q}_{\alpha}(W^*) = \{q(x,t): P(p(x,t) < q(x,t) | y) = \alpha, x \in A, t \in [T_{1}, T_{2}] \}.
\end{equation}
 Similarly, we define the \textit{exceedance probability surface} for a given threshold $l$ as 
\begin{equation}
\label{eq:exceed_surf}
\mathcal{R}_l(W^*) = \{r(x,t) = P(p(x,t) > l | y): x \in A, t \in [T_{1}, T_{2}]\}.
\end{equation} 
Values of the point-wise exceedance probability $r(x,t)$ close to 1 identify locations for which prevalence is highly likely to exceed $l$, and vice-versa. \par

In public health applications, an exceedance probability surface is a suitable predictive summary when the objective is to identify areas that may need urgent intervention because they are likely to exceed a policy-relevant prevalence threshold, say $l$. A disease ``hotspot'' is then operationally defined as the set of locations $x$, at a given time $t$, such that $p(x,l)>l$. \par

In some cases, summaries by administrative areas can be operationally useful. For example, the district-wide average prevalence for a district $D$ at time $t$ is
\begin{equation}
\label{eq:distr_average}
p_{t}(D) = \frac{1}{|D|}\int_{D} p(x,t) \: dx,
\end{equation}
where $|D|$ is its area of $D$. Incidentally, $p_{t}(D)$ can also be estimated more accurately than the point-wise prevalence $p(x,t)$, because it uses all the available information within $D$. Quantile and exceedance probability surfaces can be defined for $p_{t}(D)$ in the obvious way.

\begin{figure}[t]
\begin{center}
\includegraphics[scale=0.5]{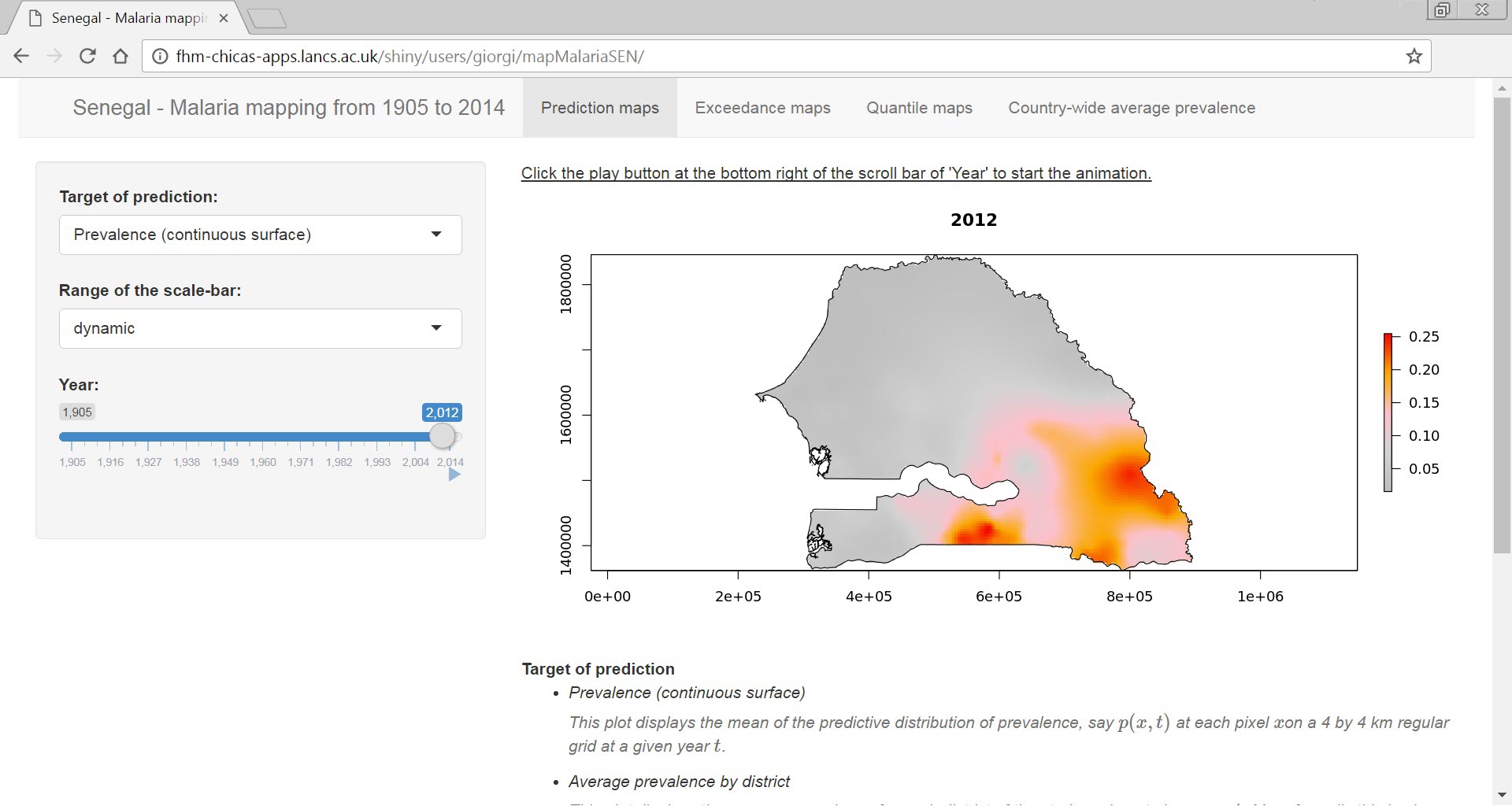}
\caption{User interface of a Shiny application for visualization of results. The underlying data are described in Section \ref{sec:app}. \label{fig:shiny_interface}}
\end{center}
\end{figure}

\subsection{Visualization}
\label{subsec:visualization}
The output from the prediction step consists of a set of $N$ predictive surfaces, whether estimates, quantiles or exceedance probabilities, within the region of interest $A$ at times $t_{1}<t_{2}<\ldots<t_{N}$. Animations then provide a useful tool for visualizing the predictive spatio-temporal surfaces and highlighting the main features of the interpolated pattern of prevalence. The R package \texttt{animation} \citep{xie2013} provides utilities for writing animations in several video and image formats. However, if interactivity is also desired, web-based ``Shiny'' applications (SAs) \citep{shiny2013} represent one of the best alternatives within R. \par
For the analysis carried out in Section \ref{sec:app}, we have developed an SA which can be viewed at
\begin{center}
{\tt http://fhm-chicas-apps.lancs.ac.uk/shiny/users/giorgi/mapMalariaSEN/}.
\end{center}

The user-interface of this SA is shown in Figure \ref{fig:shiny_interface}. Any of four panels can be chosen in order to display predictive maps of 
prevalence (``Prediction maps''), exceedance probabilities with user defined prevalence thresholds (``Exceedance maps''), quantile surfaces (``Quantile maps'') and country-wide summaries (``Country-wide average prevalence''). In the first three panels, the user can choose which target of prediction to display from a list and select the year on a slide bar. The range of prevalence and exceedance probabilities used to define the colour scale can be set to the observed range across the whole time series (``fixed'') or specific to each year (``dynamic''). The former option is convenient for comparisons between years, whilst the latter gives a more effective visualization of the spatial heterogeneity in the predictive target in a given year.

\section{Case-study: historical mapping of malaria prevalence in Senegal from 1905 to 2014}
\label{sec:app}
We analyse malaria prevalence data from 1,334 surveys conducted in Senegal between 1905 and 2014. 
The data were assembled from three different data sources: historical archives and libraries of ex-colonial institutes; online electronic databases with data on malaria infection prevalence published since the 1980s; national household sample surveys. In assembling the data for the analysis, we only included locations that were classified as individual villages or communities or a collection of communities within a definable area that does not exceed 5 km$^2$. For more details on the data extraction, see \citet{snow2015wr}. \par

\begin{figure}
\begin{center}
\includegraphics[scale=0.6]{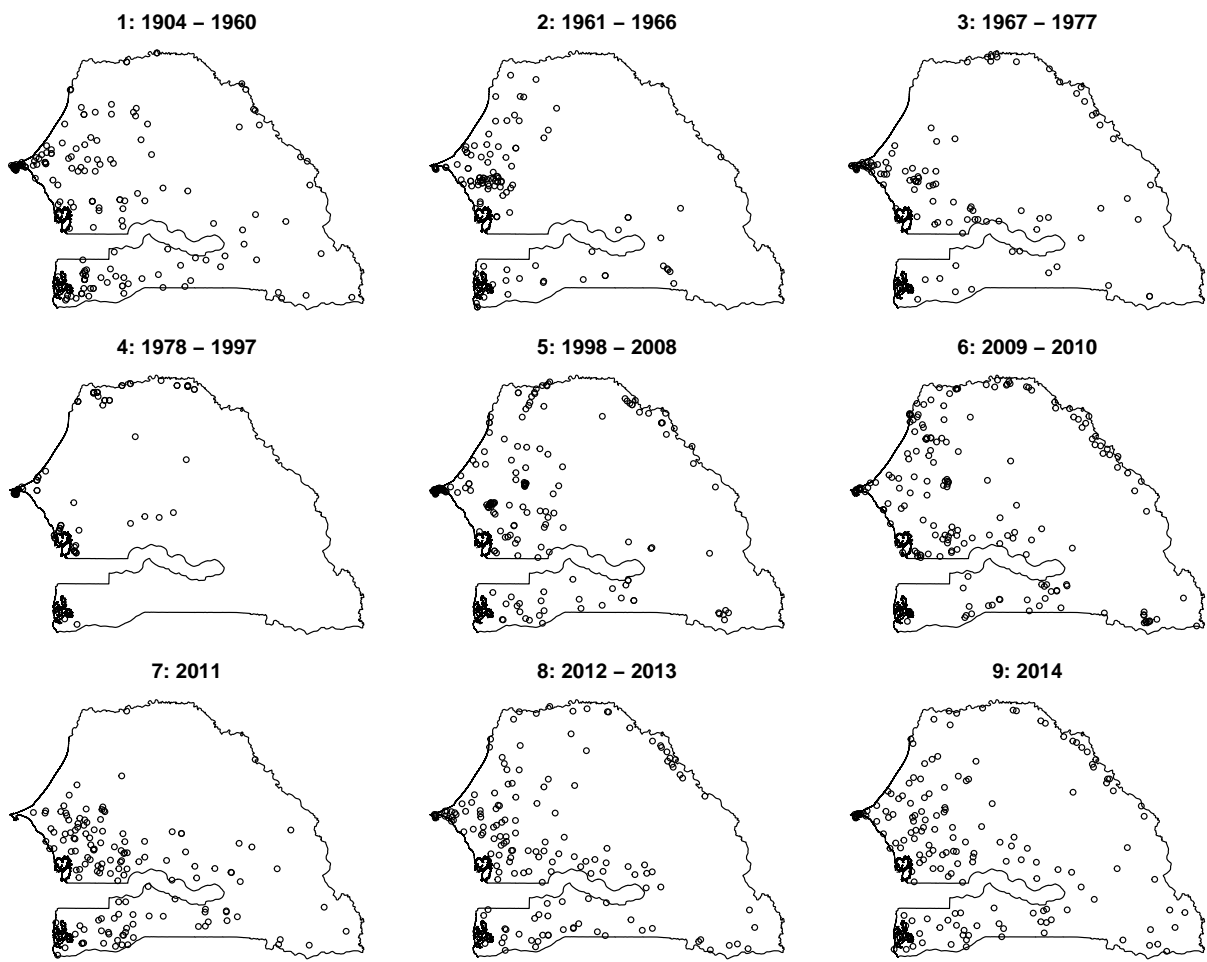}
\caption{Locations of the sampled communities in each of the time-blocks indicated by Table \ref{tab:summary_surveys}. \label{fig:sampled_loc}}
\end{center}
\end{figure}

The outcome of interest is the count $y_{i}$ of positive microscopy tests out of $n_{i}$ for {\it Plasmodium falciparum}, at a community location $x_{i}$ and year $t_{i}$.  Table \ref{tab:summary_surveys} shows the number of surveys and the average prevalence for each of the indicated time-blocks. These were identified by grouping the data points so that each time-block contains 
at least 100 surveys. We observe that 649 out of the 1334 surveys were carried out between 2009 and 2014. Also, the empirical country-wide average prevalence steadily declines from the first to the last time-block. Figure \ref{fig:sampled_loc} displays the sampled community locations within each of the time-blocks. The plot suggests a poor spatial coverage of Senegal in some years. The use of geostatistical methods can therefore be beneficial since it allows us to borrow the strength of information by exploiting the spatio-temporal correlation in the data. \par

\begin{table}[ht]
\centering
\caption{Number of surveys and country-wide average \textit{Plasmodium falciparum} prevalence, in each time-block. \label{tab:summary_surveys}}
\begin{tabular}{lcc}
  \hline
Time-block & Number of surveys & Average prevalence \\ 
  \hline
1: 1904 - 1960 & 180 & 0.416 \\ 
  2: 1961 - 1966 & 109 & 0.384 \\ 
  3: 1967 - 1977 & 104 & 0.402 \\ 
  4: 1978 - 1997 & 101 & 0.134 \\ 
  5: 1998 - 2008 & 191 & 0.111 \\ 
  6: 2009 - 2010 & 187 & 0.051 \\ 
  7: 2011 & 140 & 0.043 \\ 
  8: 2012 - 2013 & 157 & 0.038 \\ 
  9: 2014 & 165 & 0.019 \\ 
   \hline
\end{tabular}
\end{table}

Our model for the data is of the form \eqref{eq:std_model}, with the following linear predictor 
\begin{eqnarray}
\label{eq:std_model}
\log\left\{\frac{p(x_{i},t_{i})}{1-p(x_{i},t_{i})}\right\} &=& \beta_{1}+\beta_{2}a(x_{i}, t_{i}) + \beta_{3}[a(x_{i},t_{i})-5]\times I\{a(x_{i},t_{i})>5\} + \nonumber \\ && 
\beta_{4}A(x_{i}, t_{i}) + \beta_{5}[A(x_{i},t_{i})-20]\times I\{A(x_{i},t_{i})>20\} + \nonumber \\ 
&& 
S(x_{i}, t_{i})+Z(x_{i}, t_{i}),
\end{eqnarray}
where $a(x_{i},t_{i})$ and $A(x_{i}, t_{i})$ are the lowest and largest observed ages among the sampled individuals at location $x_{i}$ and time $t_{i}$, respectively. In \eqref{eq:std_model}, we use linear splines, each with a single knot, at 5 years for $a(x,t)$ and at 20 years for $A(x,t)$. For the spatio-temporal process $S(x,t)$, we use a Gneiting correlation function, as in \eqref{eq:gneiting_cor1}, with $\delta=\xi=0$, i.e. a separable covariance function. \par

\begin{figure}
\begin{center}
\includegraphics[scale=0.44]{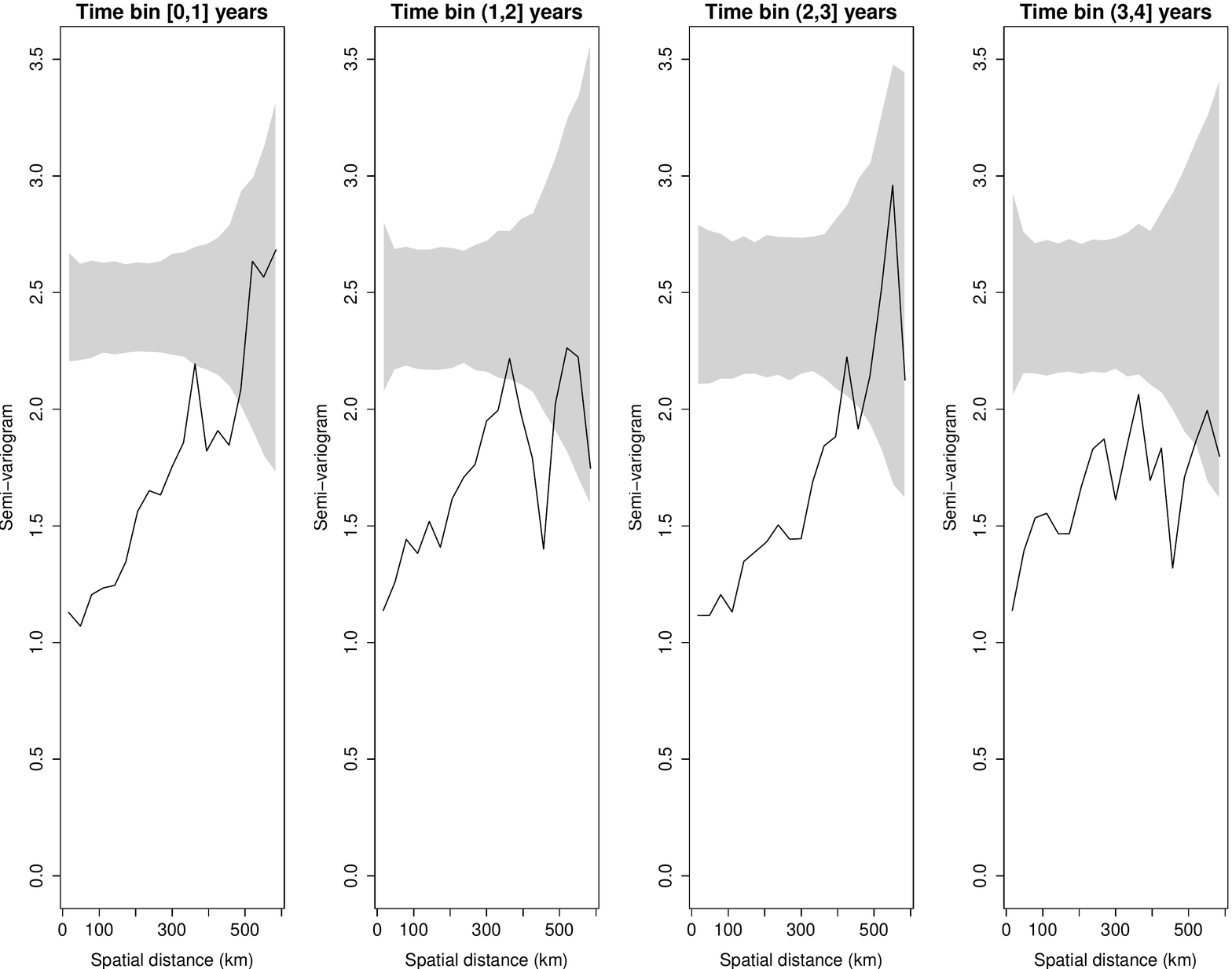}\\
\vspace{0.5cm}
\includegraphics[scale=0.44]{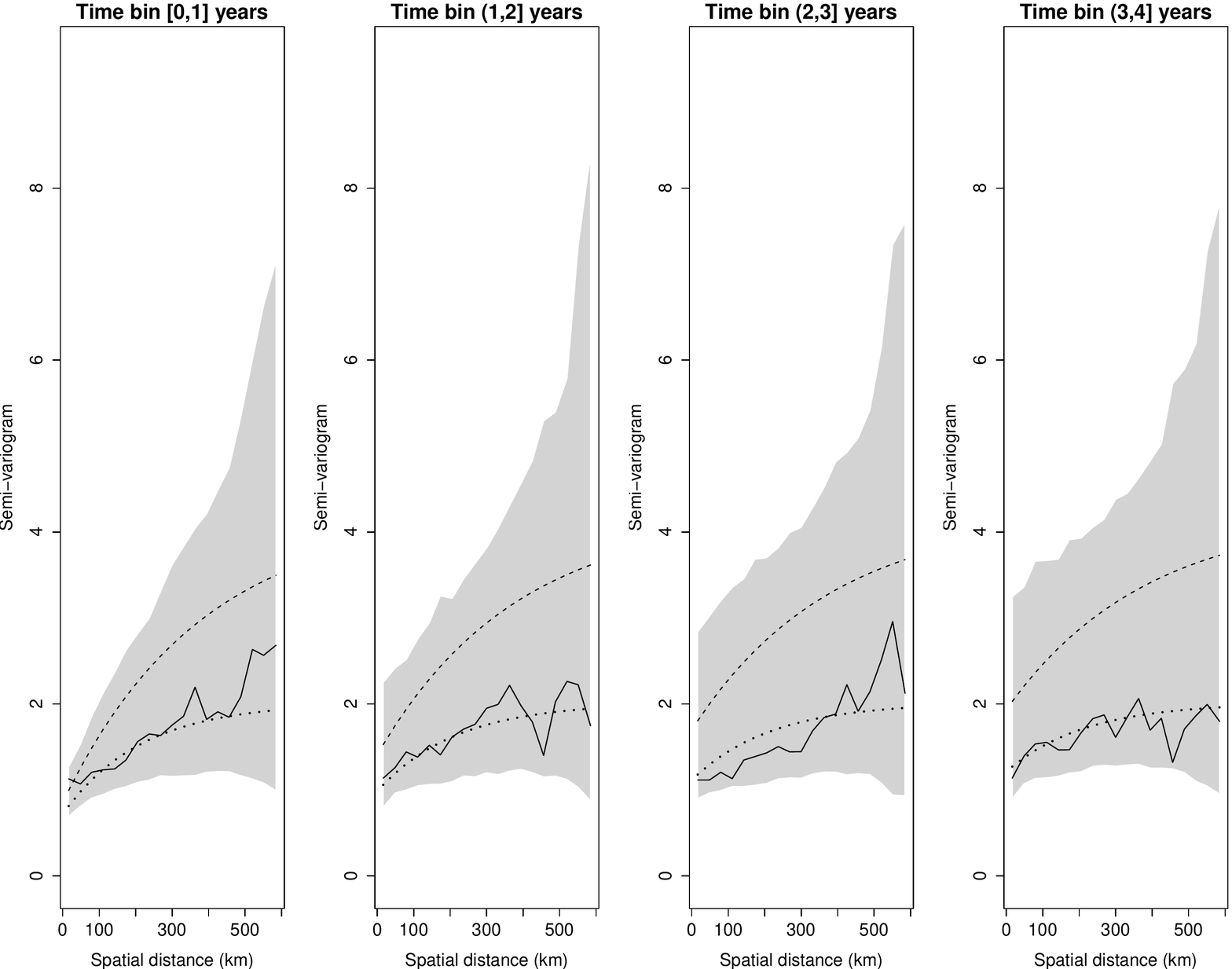}
\caption{The plots show the results from the Monte Carlo methods used to test the hypotheses of spatio-temporal indepence (upper panels) and of compatibility of the adopted covariance model with the data (lower panels). The shaded areas represent the 95$\%$ tolerance region under each of the two hypotheses. The solid lines correspond to the empirical variogram for $\tilde{Z}(x_{i}, t_{i})$, as defined in Section \ref{subsec:expl_analysis}.  In the lower panels, the theoretical variograms obtained from the least squares (dotted lines) and maximum likelihood (dashed lines) methods are shown. \label{fig:vario_diag}}
\end{center}
\end{figure}

Using the predictive mean as a point estimate of the random effects from a non-spatial binomial mixed model, we carry out the test for residual spatio-temporal correlation, as outlined in Section \ref{subsec:expl_analysis}. The upper panels of Figure \ref{fig:vario_diag} show overwhelming evidence against the assumption of spatio-temporal independence. We then initialize the covariance parameters, $\phi$ and $\psi$, using a least squares fit to the empirical variogram, as shown by the dotted lines in the lower panels of Figure \ref{fig:vario_diag}.  \par

\begin{table}[ht]
\centering
\caption{Maximum likelihood estimates of the model parameters and their $95\%$ confidence intervals (CI) based on the asymptotic Gaussian approximation (GA) and parametric bootstrap (PB). \label{tab:mle}}
\begin{tabular}{rrcc}
  \hline
 Parameter & Estimate & $95\%$ CI (GA) & $95\%$ CI (PB) \\ 
  \hline
$\beta_{1}$ & -1.830 & (-3.180, -0.480) & (-3.131, -0.367) \\ 
$\beta_{2}$ & 0.118 & (0.017, 0.220) & (0.019, 0.226) \\ 
$\beta_{3}$ & -0.334 & (-0.562, -0.105) & (-0.585, -0.103) \\ 
$\beta_{4}$ & 0.015 & (-0.022, 0.052) & (-0.025, 0.052) \\ 
$\beta_{5}$  & -0.014 & (-0.055, 0.027) & (-0.056, 0.030) \\ 
$\sigma^2$ & 3.650 & (2.378, 5.601) & (2.272, 5.222) \\ 
$\phi$ & 381.022 & (225.948, 642.528) & (220.593, 568.953) \\ 
$\tau^2/\sigma^2$ & 0.157 & (0.097, 0.253) & (0.105, 0.253) \\ 
$\psi$ & 6.730 & (3.571, 12.683) & (3.484, 10.669) \\ 
   \hline
\end{tabular}
\end{table}

\begin{table}[ht]
\centering
\caption{Posterior mean and 95$\%$ credible intervals of the model parameters from the Bayesian fit. \label{tab:bayes}}
\begin{tabular}{rcc}
  \hline
 & Posterior mean & $95\%$ credible interval \\ 
  \hline
$\beta_{1}$ & -1.899 & (-3.746, -0.275) \\ 
$\beta_{2}$ & 0.116 & (0.013, 0.212) \\ 
$\beta_{3}$ & -0.335 & (-0.560, -0.115) \\ 
$\beta_{4}$ & 0.013 & (-0.023, 0.050) \\ 
$\beta_{5}$ & -0.013 & (-0.054, 0.028) \\ 
$\sigma^2$ & 4.649 & (2.887, 7.641) \\ 
$\phi$ & 504.330 & (283.019, 863.198) \\ 
$\tau^2/\sigma^2$ & 0.137 & (0.075, 0.217) \\ 
$\psi$ & 9.098 & (4.443, 16.608) \\ 
   \hline
\end{tabular}
\end{table}

\begin{figure}
\begin{center}
\includegraphics[scale=0.42]{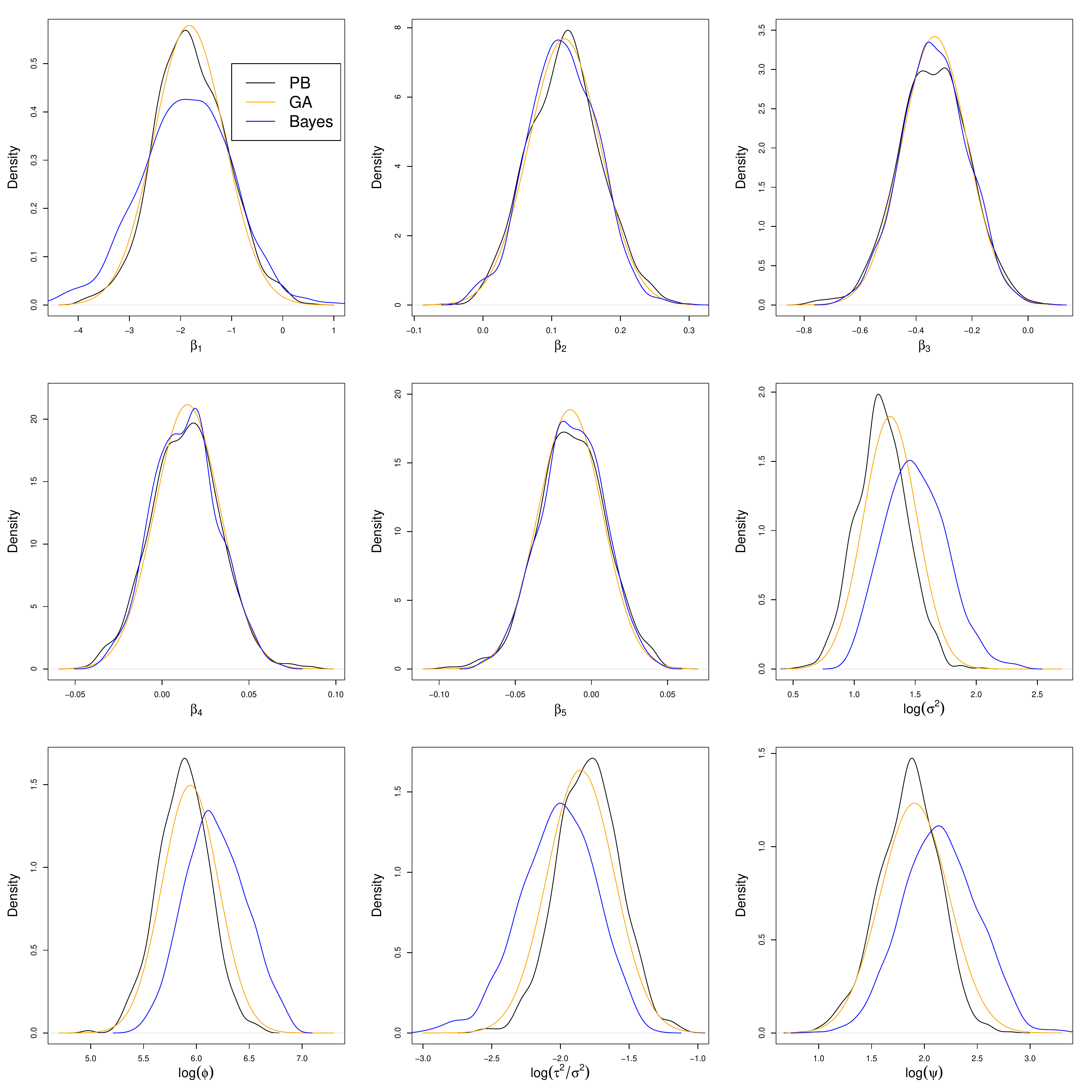}
\caption{Density functions of the maximum likelihood estimator for each of the model parameters based on parameteric bootstrap (PB), as black lines, and the Gaussian approximation (GA), as orange lines; the blue lines correspond to the posterior density from the Bayesian fit. \label{fig:param_uncertainty}}
\end{center}
\end{figure}

We conducted parameter estimation and spatial prediction using both likelihood-based and Bayesian inference. In the latter case, we specifed the following set of independent and vague priors: $\beta \sim MVN(0,10^4 I)$; $\sigma^2 \sim \text{Uniform}(0,20)$; $\phi \sim \text{Uniform}(0,1000)$; $\tau^2/\sigma^2 \sim \text{Uniform}(0,20)$; $\psi \sim \text{Uniform}(0,20)$.
Table \ref{tab:mle} shows the maximum likelihood estimates of the model parameters and their corresponding 95\% confidence intervals
based on the Gaussian approximation (GA) and on parametric boostrap (PB),
together with Bayesian esimates (posterior means) and 95\% credible intervals.
 The two non-Bayesian methods
 give similar confidence intervals; the difference is noticeable, although still small in practical terms,
  only for the parameter $\phi$. The Bayesian method gives materially larger estimates of $\sigma^2$ and
 $\phi$ .  Note that for both of these parameters,  the prior means are substantially larger than the
maximum likelihood estimates, suggesting that the priors, although vague, have 
nevertheless had some impact on the estimates.

 Figure \ref{fig:param_uncertainty} gives a different perspective on the similarities and differences between the
results obtained by 
the non-Bayesian and Bayesian methods.  The Bayesian
posterior density of the intercept has heavier tails than the
sampling distribution of the maximimum likelihood estimator;  the posterior densities of $\sigma^2$, $\phi$ and $\psi$ 
are shifted to the right of their non-Bayesian counterparts,
whilst the posterior density of $\tau^2/\sigma^2$ is shifted to the left. Finally, there is some
residual skewness in the PB distributions of the log-transformed covariance parameters. \par

Using the Monte Carlo methods of Section \ref{subsec:diagnostic}, we checked the validity of the assumed covariance model. The lower panels of Figure \ref{fig:vario_diag}
show that 
 for each of the four time-lag intervals considered, the observed variograms fall within
within the 95$\%$ tolerance region obtained under the fitted model; the $p$-value for a 
Monte Carlo goodness-of fit
test  using the test statistic \eqref{eq:test_statistic} is 0.548.


\begin{figure}
\begin{center}
\includegraphics[scale=0.7]{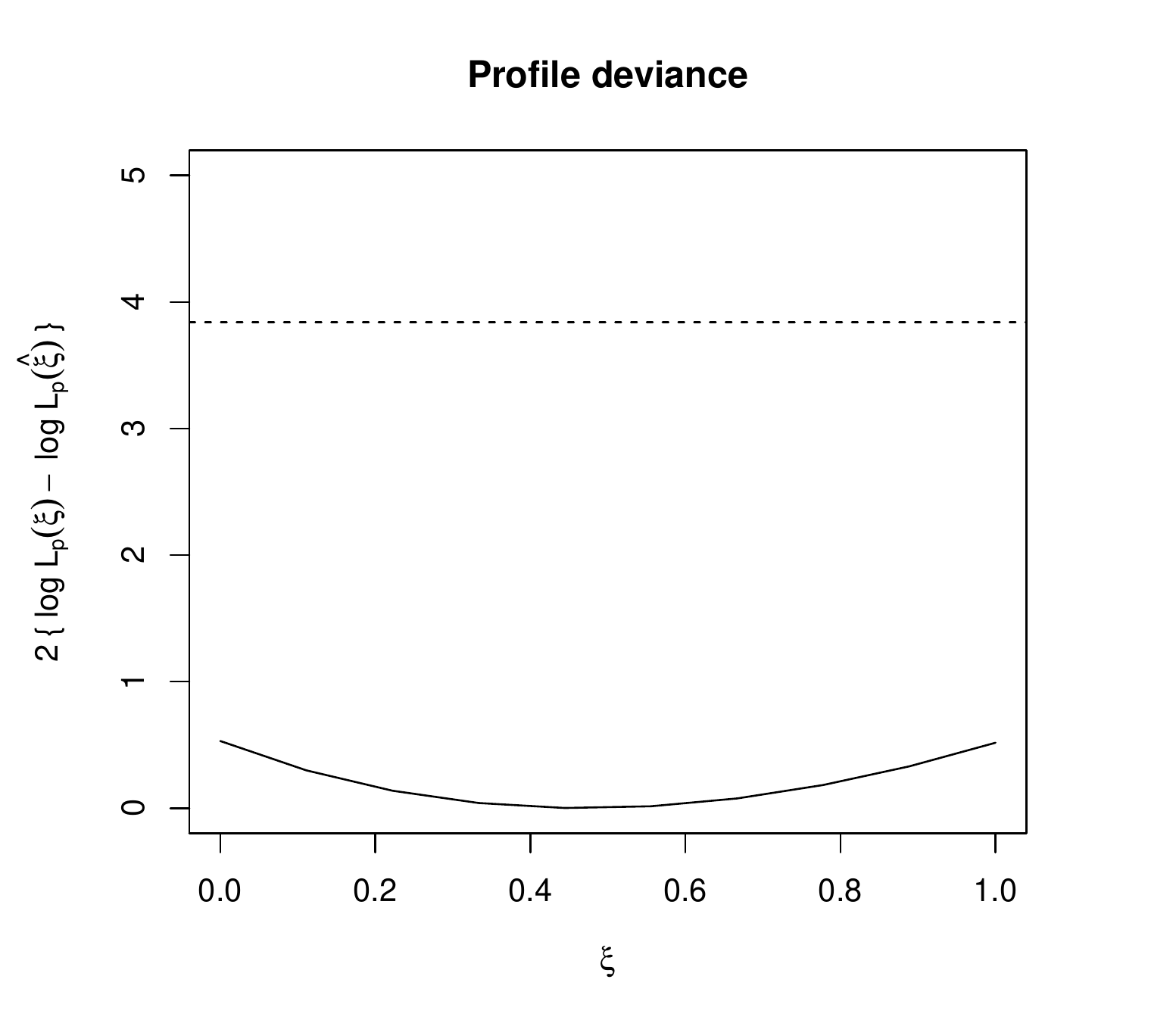}
\caption{Profile deviance (solid line) for the parameter of spatio-temporal interaction $\xi$ of the \citet{gneiting2002} family given by \eqref{eq:gneiting_cor1}. The dashed line is the 0.95 quantile of a $\chi^2$ distribution with one degree of freedom. \label{fig:profile_xi}}
\end{center}
\end{figure}

\begin{figure}
\begin{center}
\includegraphics[scale=0.9]{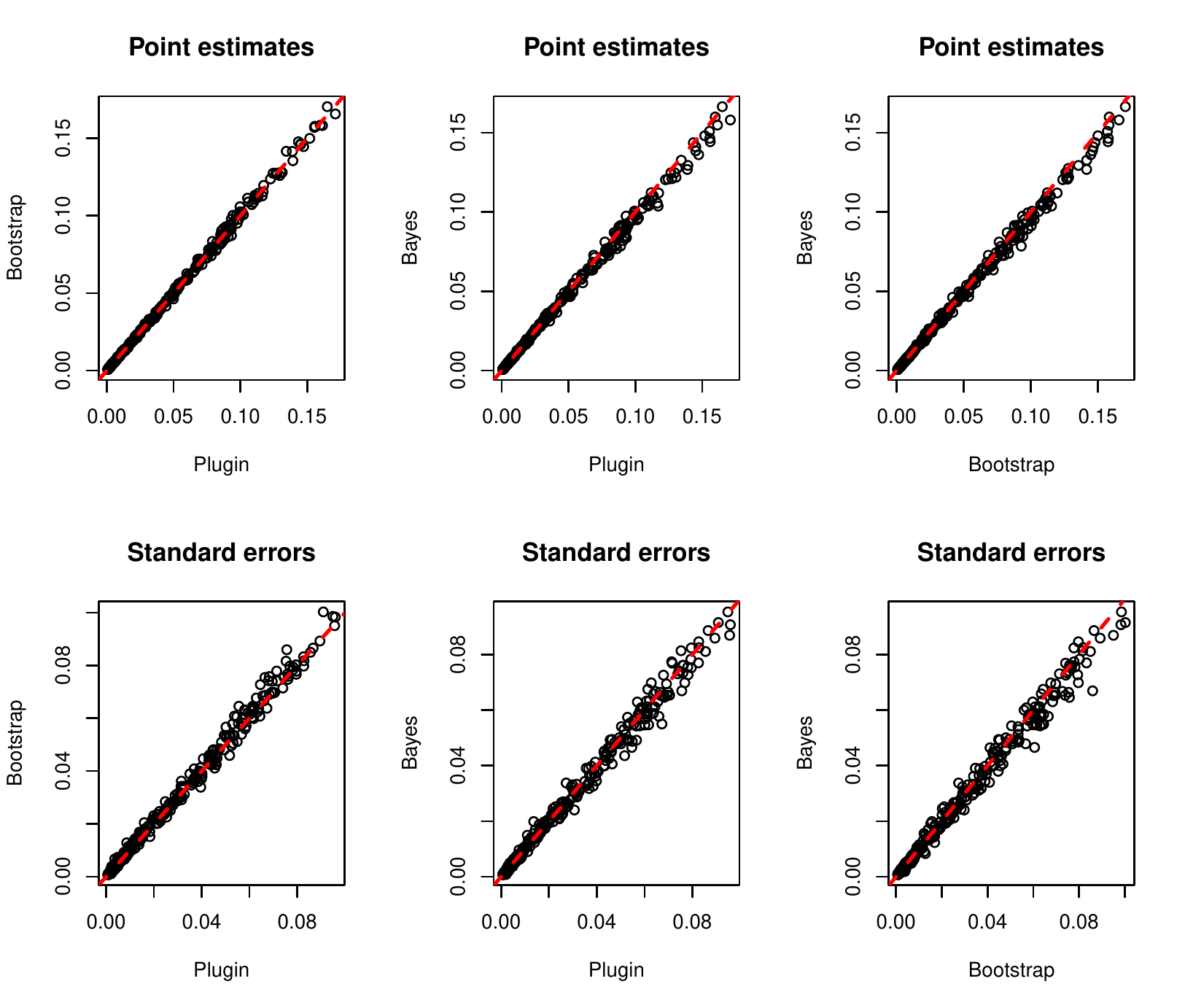}
\caption{Scatter plots of the point estimates (upper panels) and standard errors (lower panels) of \textit{Plasmodium falciparum} prevalence for children between 2 and 10 years of age, using plugin, parametric bootsptrap and Bayesian methods. The dashed red lines in each panel is the identity line. \label{fig:compare_pred}}
\end{center}
\end{figure}

Figure \ref{fig:profile_xi} shows the profile deviance  function
$$
D(\xi) = 2\{\log L_p(\hat{\xi})-\log L_{p}(\xi)\},
$$
where $L_{p}(\xi)$ is the profile likelihood for the parameter of spatio-temporal interaction parameter $\xi$ and $\hat{\xi}$ is its Monte Carlo maximum likelihood estimate. The dashed horizontal line is the 0.95 quantile of a $\chi^2$ distribution with one degree of freedom. The flatness of $D(\xi)$ indicates that data give very little information
 about the non-separability of the correlation structure of $S(x,t)$.

\begin{figure}
\begin{center}
\includegraphics[scale=0.6]{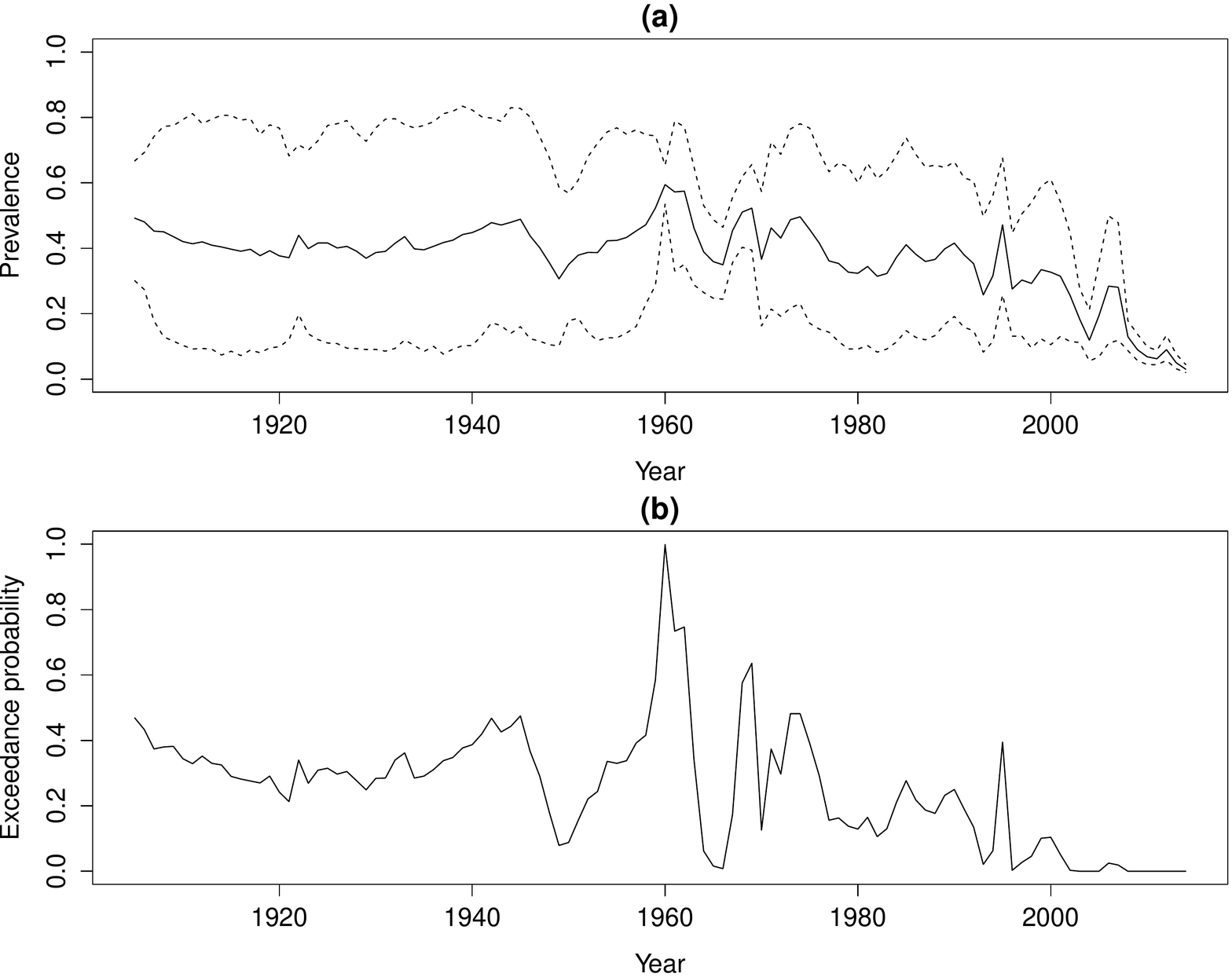}
\end{center}
\caption{(a) Predictive mean (solid line) of the country-wide average prevalence with 95$\%$ predictive intervals. (b) Predictive probability of the country-wide average prevalence exceeding a 50$\%$ threshold.  \label{fig:estim}}
\end{figure}

\begin{figure}
\begin{center}
\includegraphics[scale=0.85]{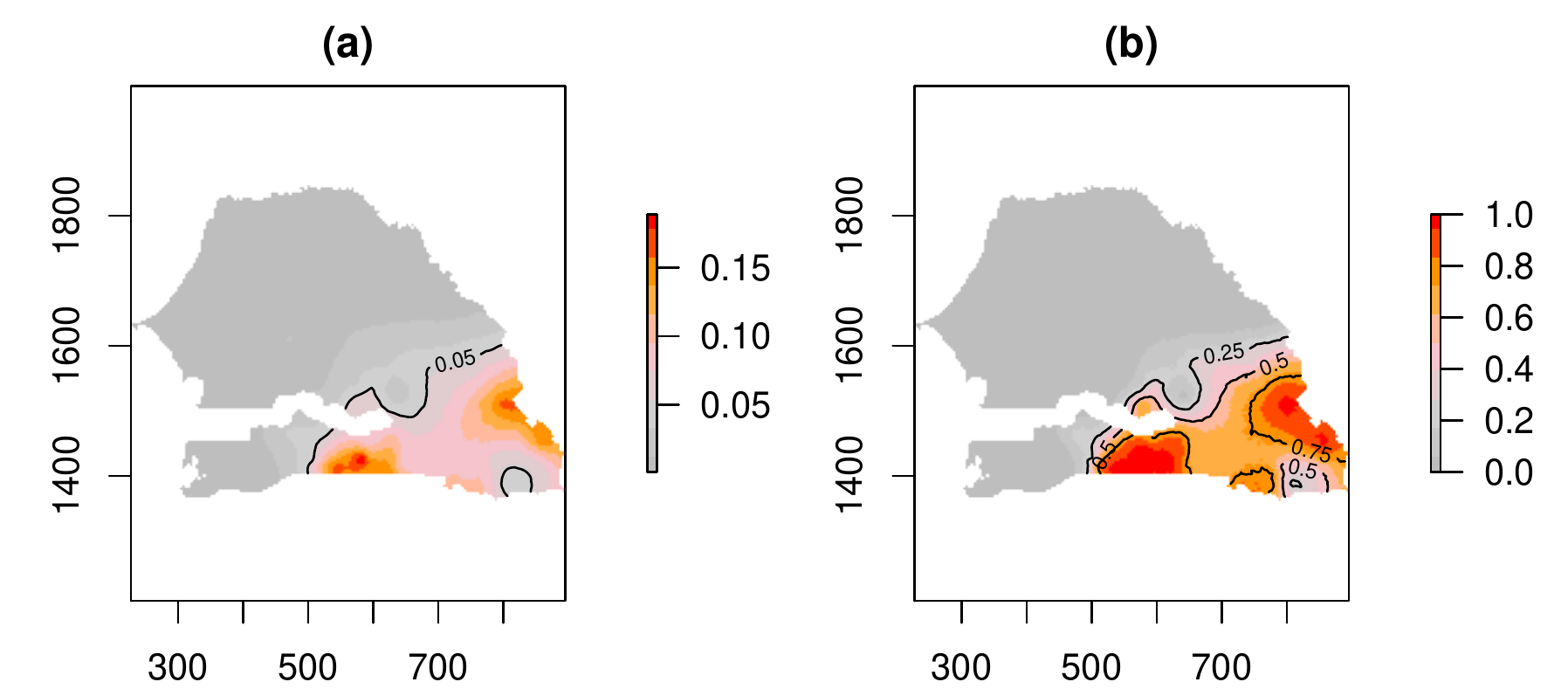}
\caption{(a) Predictive mean surface of prevalence for children between 2 and 10 ($PfPR_{2-10}$); (b) Exceedance probability surface for a threshold of $5\%$ $PfPR_{2-10}$. Both maps are for the year 2014. The contour lines correspond to 5$\%$ $PfPR_{2-10}$, in the left panel, and to 25$\%$, 50$\%$ and 75$\%$ exceedance probability, in the right panel. \label{fig:prev_exceed}}
\end{center}
\end{figure}

To assess the differences in the spatial predictions 
obtained using thr GA, PB and  Bayesian approaches, we used each method to
predict \textit{P. falciparum} prevalence for children between 2 and 10 years of age ($PfPR_{2-10}$)
 in the year 2014, at each point on a 10 by 10 km regular grid covering the whole of Senegal. Figure \ref{fig:compare_pred} shows pairwise scatterplots of the three sets of
point predictions and associated standard deviations of $PfPR_{2-10}$. All six scatter plots show only small
deviations from the identity line.\par

Figure \ref{fig:estim}(a) shows point and interval predictions of  average country-wide  $PfPR_{2-10}$. 
We observe a steady decline in $PfPR_{2-10}$ in the most recent decade. The highest
predicted value of $PfPR_{2-10}$ across the whole of the time series occured in 1960,
the year in which Senegal gained independence from France.  Figure \ref{fig:estim}(b)  
shows for each year the predictive probability that average country-wide
 $PfPR_{2-10}$ exceeded 5\%. 
 Figure \ref{fig:prev_exceed} shows the surfaces of the predictive mean (left panel) and the preditive probability
that prevalence exceeds
 $5\%$ prevalence (right panel), for the year 2014. In the right panel, we can identify two disjoint areas in the south-west of Senegal, where the probability of exceeding $5\%$ $PfPR_{2-10}$ is at least 75$\%$. In areas between the contour of 50$\%$ and 75$\%$ exceedance probability we are less 
confident that $PfPR_{2-10}$ exceeds $5\%$. These aspects relating to the uncertainty about the $5\%$ threshold cannot be deduced from the map of prevalence estimates in the left panel, nor would a map of pointwise prediction
variances be of much help.
 
\section{Discussion}
\label{sec:discussion}
We have developed a statistical framework for the analysis of spatio-temporally referenced data from repeated cross-sectional
prevalence surveys. Our aim was to provide a set of tools and principles that can be used to identify a parsimonious geostatistical model that is compatible with the data. In our view, model validation should include checking the validity of the specific assumptions made on $S(x,t)$ rather than be focused exclusively on predictive performance, so as to avoid the risk of attaching spurious precision to predictions from an inappropriate model. \par

The variogram is very widely used in geostatistical analysis.  We use it
 both for exploratory analysis and model validation, but favour likelihood-based methods,
whether non-Bayesian or Bayesian, for parameter estimation and formal model comparison; an example of
the latter is our use of  the profile deviance to justify fitting a model with separable correlation
structure to the Senegal malaria data. \par

In our spatio-temporal analysis of historical malaria prevalence data from Senegal, we have shown how to incorporate parameter uncertainty within a likelihood-based framework by approximation of the distribution of the maximum likelihood estimator using the Gaussian approximation and parametric bootstrap. The results showed that the Gaussian approximation provides reliable numerical inferences for the regression coefficients but was slightly inaccurate for the log-transformed covariance parameters. For this reason, we generally recommend using parametric bootstrap whenever this is computationally feasible. In our view, this gives a viable approach to handling parameter uncertainty in predictive inference without requiring the specification of so-called non-informative priors. Non-Bayesian
and  Bayesian approaches showed some differences with respect to parameter
estimation, but delivered almost identical point predictions and predictive standard deviations for the spatial estimates of prevalence. Our results also illustrate how even large geostatistical data-sets often lead to 
disappointingly imprecise inferences about model parameters.  For this reason, 
we woild favour Bayesian inference when, and only when,  an informative prior can be specified from
contextually based  expert prior knowledge of the process under investigation. \par

In Section \ref{subsec:diagnostic}, we  discussed how to extend the standard model for prevalence data in order to let the model parameters change over time, space or both. However, the use of these models requires a large amount of the data and good spatio-temporal coverage so as to detect non-stationary patterns in prevalence. In the Senegal malaria
application application  the spatio-temporal sparsity of the sampled locations meant that
 the data could not be used to reliably detect spatio-temporal variation in the covariance parameters. For this
application 
we also assumed that the sampling locations did not arise from a preferential sampling scheme. The standard geostatistical model for prevalence can also be extended to account for preferentiality in the sampling design, based on the framework developed by \citet{diggle2010}. However, such a model would require a larger amount of data than was available for this application. \par

Our analysis included data from the Demographic and Health Survey (DHS) conducted in Senegal in 2014. These data were collected using a two-stage stratified sampling design \citep{dhs-senegal2014}. In the first stage, 200 census districts (CDs) are randomly selected, 79 among urban CDs and 121 among rural CDs, with probability proportional to the population size. In
the second stage, an enumeration list from each CD was used to sample households randomly. In the analysis reported above,
 we could not account for the sampling design of the DHS data because of the lack of information on urban and rural extents for every single year when the surveys were conducted. However, since this variable is available for 2014, we extracted the DHS data and fitted two geostatistical models with and without an explanatory variable that classifies every location as rural or urban. Figure \ref{fig:urban_rural} shows the plots for the estimated prevalence and associated standard errors obtained from the two models. The differences both in the point estimates and standard error of prevalence are negligible. Hence, we do not expect the sampling design adopted in the DHS survey to affect the results reported in Section \ref{sec:app}.

\begin{figure}
\begin{center}
\includegraphics[scale=0.8]{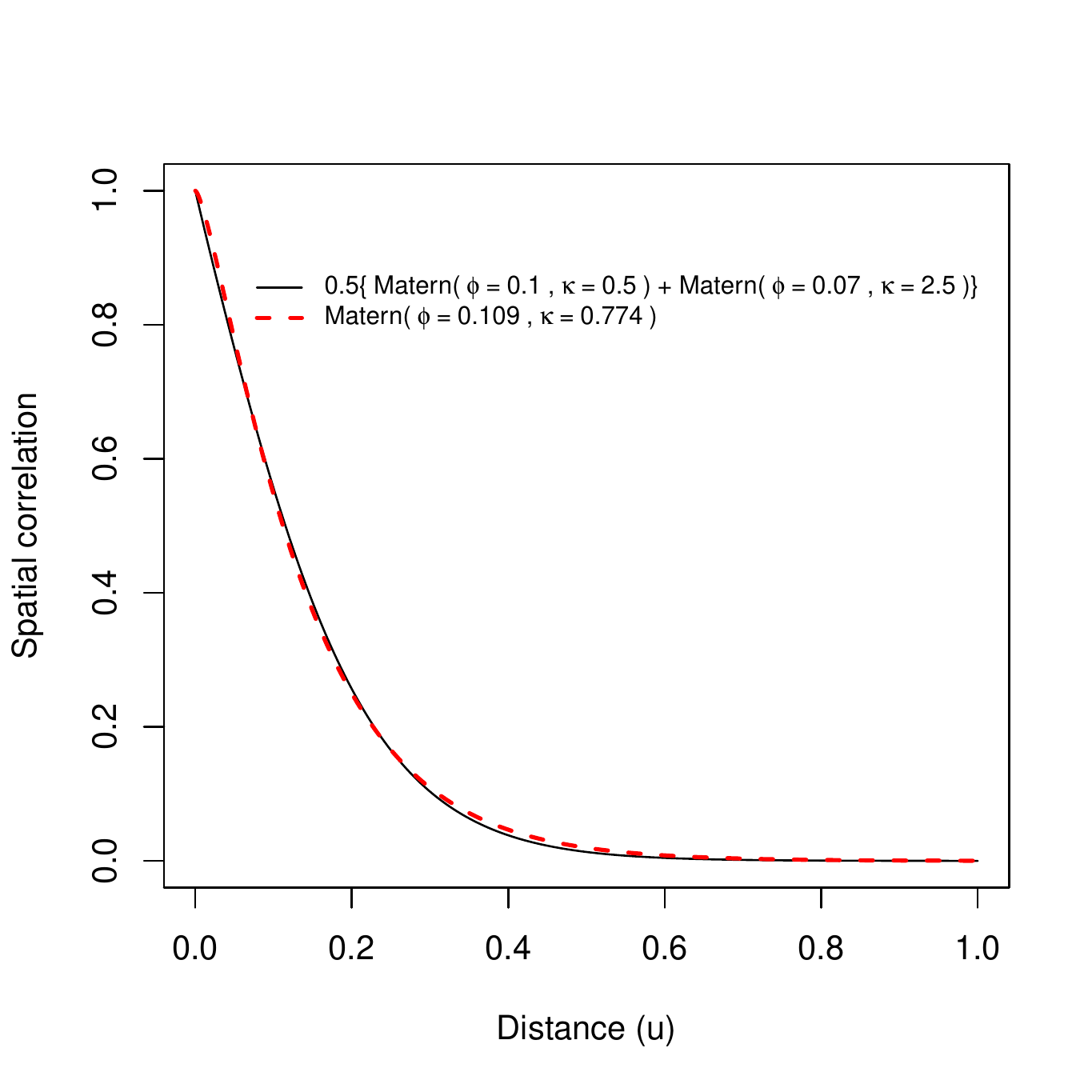}
\caption{The solid curve corresponds to the function $f_{1}(u) = 0.5\{\mathcal{M}(u;0.1,0.5)+ \mathcal{M}(u;0.07,2.5)\}$ and the red dashed curve to $\mathcal{M}(u;0.109,0.774)$, where $\mathcal{M}(\cdot; \phi, \kappa)$ is a Mat\'ern correlation function with scale parameter $\phi$ and smoothness parameter $\kappa$. \label{fig:corr_function}}
\end{center}
\end{figure}

\begin{figure}
\begin{center}
\includegraphics[scale=0.65]{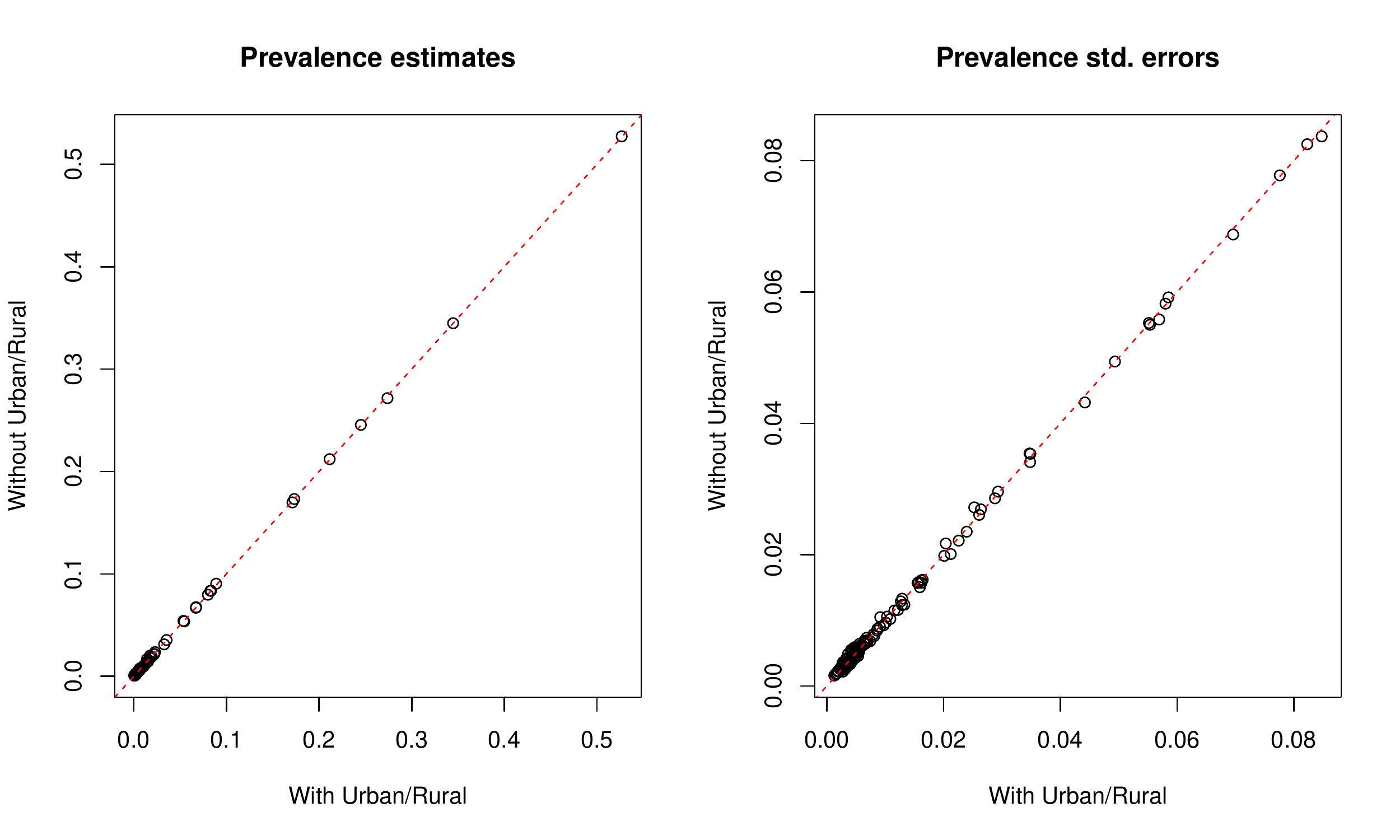}
\caption{Prevalence estimates (left panel) and standard errors (right panel) based on the Demographic and Health Survey conducted in Senegal in 2014. Those are obtained from a model using a spatial indicator for urban and rural communities (x-axis) and excluding this explanatory variable (y-axis). The dashed line in both graphs is the identity line. \label{fig:urban_rural}}
\end{center}
\end{figure}

In model \eqref{eq:st_model}, spatial confounding  can
 arise when some of the variation
in prevalence due to to the
effect of spatially structured risk factors $d(x,t)$ is attributed by the model to the stochastic process $S(x,t)$.
This phenomenon affects the interpretation of 
the regression parameters $\beta$;  see, for example, \citet{paciorek2010} and \citet{hodges2010}. 
 However, the following argument supports our experience
that it has a negligible impact on predictive inference for $p(x,t)$. Consider, for simplicity, the
following purely spatial model,
\begin{equation}
\log\left\{\frac{p(x_{i})}{1-p(x_{i})}\right\} = \beta_{0}+\beta_{1}D_{1}(x_i)+\beta_{2}D_{2}(x_i) + S(x_i).
\label{eq:true_model}
\end{equation}
 If both
of $D_1(x)$ and $D_2(x)$ are observed, fitting the model (\ref{eq:true_model}) with $D_1(x)$
and $D_2(x)$ as covariates, i.e.  conditioning on
both $D_1(x)$ and $D_2(x)$, would lead to consistent estimation of $\beta_1$ and $\beta_2$. If only
$D_1(x)$ is observed, we can only condition on $D_1(x)$.  Now  assume that $D_{2}(x) = T(x) + D_{1}(x)$, with $S(x)$ and  $T(x)$ independent processes, and re-express
 (\ref{eq:true_model})  as
\begin{eqnarray}
\log\left\{\frac{p(x_{i})}{1-p(x_{i})}\right\} & = & \beta_{0}+\beta_{1} D_1(x_i) +\beta_{2}\{T(x_i)+D_{1}(x_i)\} + S(x_i) + Z(x_i) \nonumber \\
& = & \beta_{0}+ \beta_{1}^* D_1(x_i) + S^*(x_i)
\label{eq:re-expression}
\end{eqnarray}
where $\beta_1^* = \beta_1 + \beta_2$ and $S^*(x) = S(x) + \beta_{2}T(x)$. 
Provided that we correctly specify the model for $S^*(x)$,
conditioning on $D_1(x)$ will lead to consistent estimation of $\beta^*$,
which is  all that we require for prediction of $p(x)$.  Now suppose that
 $T(x)$ and $S(x)$ are Mat\'ern processes, but  we  specify $S^*(x)$ to be
a Mat\'ern process.  This is incorrect, but we conjecture that it is a good approximation.
 Figure \ref{fig:corr_function} shows an example in which $\beta_{2}=1$ and  $S(x)$ and $T(x)$ 
have Mat\'ern covariance functions with unit variance, scale parameters $0.1$ and $0.07$ and smoothness parameters $0.5$ and $2.5$, respectively. The
resulting  correlation  function of $S^*(x)$ is $f_{1}(u) = 0.5\{\mathcal{M}(u;0.1,0.5)+ \mathcal{M}(u;0.07,2.5)\}$, which can be closely
approximated by a single Mat\'ern, $f_{2}(u) = \mathcal{M}(u;0.109,0.774)$, where $\mathcal{M}(\cdot; \phi, \kappa)$ is a Mat\'ern correlation function with scale parameter $\phi$ and smoothness parameter $\kappa$.  

For large data-sets, it may be necessary to use an
approximation of the spatio-temporal Gaussian process $S(x,t)$ in order to make inference computationally feasible. 
One such approach is to use a low-rank approximation   \citep{higdon1998,higdon2002} in which 
$S(x,t)$ is represented as a finite linear combination of basis functions  with radom coefficients; see, for example, \citet{rodrigues2010} who develop a class of non-separable spatio-temporal covariance functions using 
this approach.  Another approach is to formulate
$S(x,t)$ as the solution to a stochastic partial differential equation (SPDE). 
\citet{lindgren2011} develop a general framework for this approach, in which Gaussian Markov random fields
are used to obtain a computationally fast solution to a discretised version of the defining SPDE.
In the case of binary data, the computational burden can also be
reduced by using data augmentation sampling schemes \citep{holmes2006}. \par

Throughout the paper, we have assumed that the process $S(x,t)$ is isotropic. To diagnose anisotropy, a directional
version of the variogram
can be used, in which
 inter-point distances $u$ are replaced by vector differences $x_{i}-x_{j}$ and the results displayed as a three-dimensional scatterplot at each time-lag. \citet{weller2016} provides a comprehensive survey of non-parametric diagnostic methods used to test specific deviations from the assumption of isotropy. A limitation of most of these methods is that they require the spatial process to be observed either on a grid or on a relaisation of 
a homogeneous Poisson process. Additionally, the properties of these tests have only been investigated when the response is continuous. The sample size required to obtain adequate power is likely to be higher in the case of binomial data. \par

In addition to the sampling designs that we discussed in Section \ref{sec:design}, cluster sampling is another cost-effective alternative to simple random sampling. In households surveys, a cluster might correspond to a geographically restricted area, e.g. a village or group of households, which are randomly selected in a first stage. One of the potential, but still unexplored, uses of this sampling design in disease mapping would be to disentangle the long-range and small-range spatial variation in disease risk. To pursue this objective the nugget component $Z(x_{i}, t_{i})$ in \eqref{eq:st_model} could be modelled as an additional Gaussian process whose scale of spatial correlation is constrained to be smaller than that of 
$S(x_{i}, t_{i})$. Separating these two spatial scales of correlation
 would require a large amount of data and would be dependent on the spatial arrangement of the clusters. \par 

We have not considered issues of data-quality variation across multiple surveys. This has been
addressed by \citep{giorgi2015}, who developed a multivariate geostatistical model to combine
prevalence data from multiple randomised and non-randomised surveys. Incorporation of
this modelling framework into the methods of Section \ref{sec:methods} would be straightforward given the
required data, since all the different stages of the analysis can still be carried out using the
same tools and principles. 

\section*{Acknowledgements}
EG holds an MRC Strategic Skills Fellowship in Biostatistics (MR/M015297/1). RWS is funded as a Principal Fellow by the Wellcome Trust, UK (No. 079080 and 103602) and is grateful to the UK’s Department for International Development for their continued support to the project \textit{Strengthening the Use of Data for Malaria Decision Making in Africa first}, funded and piloted in 2013 (DFID Programme Code No. 203155). AMN acknowledges support from the Wellcome Trust as an Intermediary Fellow (No. 095127). 

\bibliographystyle{biometrika}
\bibliography{biblio}
\end{document}